\documentclass[10pt]{amsart}
\usepackage{latexsym,amssymb}
\usepackage{graphicx}

\theoremstyle{plain}
\newtheorem{lemma}{Lemma}
\newtheorem{proposition}{Proposition}
\newtheorem{theorem}{Theorem}
\newtheorem{corollary}{Corollary}
\newtheorem{example}{Example}[section]

\addtocounter{section}{-1}

\def\cal#1{{\mathcal{#1}}}

\def\BP{\noindent {\it Proof.} }
\def\EP{\hfill$\Box$

}
\def\BC{\begin{center}}
\def\EC{\end{center}}
\def\BT{\begin{theorem}}
\def\ET{\end{theorem}}
\def\BPR{\begin{proposition}}
\def\EPR{\end{proposition}}
\def\BL{\begin{lemma}}
\def\EL{\end{lemma}}
\def\BCO{\begin{corollary}}
\def\ECO{\end{corollary}}
\def\BPI{\BC \begin{picture}}
\def\EPI{\end{picture}\EC}
\def\BTA{\begin{tabular}}
\def\ETA{\end{tabular}}
\def\BX{\begin{example}}
\def\EX{\end{example}}
\def\BIT{\begin{itemize}}
\def\EIT{\end{itemize}}
\def\IT{\item}

\def\B{{\cal B}}

\def\F{{\cal F}}

\def\I{{\cal I}}

\def\O{{\cal O}}

\def\Q{{\cal Q}}

\def\S{{\cal S}}
\def\T{{\cal T}}

\def\Y{{\cal Y}}

\def\Ro{R}

\begin{document}

\author{Marcel Ern\'{e}}
\address{Leibniz University, Faculty for Mathematics and Physics\\
Welfengarten 1, 30167 Hannover, Germany}
\email{erne@math.uni-hannover.de}
\urladdr{http://www2.iazd.uni-hannover.de/~erne/}

\title[Core spaces, sector spaces and fan spaces]
{Core spaces, sector spaces and fan spaces:\\ a topological approach to domain theory}

\date{Mai 29, 2016}

\begin{abstract}
We present old and new characterizations of core spaces, alias worldwide web spaces, 
originally defined by the existence of supercompact neighborhood bases.
The patch spaces of core spaces, obtained by joining the original topology with a second topology having the dual specialization order,
are the so-called sector spaces, which have good convexity and separation properties and determine the original space. 
The category of core spaces is shown to be concretely isomorphic to the category of fan spaces; these are certain
quasi-ordered spaces having neighborhood bases of so-called fans, 
obtained by deleting a finite number of principal filters from a principal filter. 
This approach has useful consequences for domain theory. 
In fact, endowed with the Scott topology, the continuous domains are nothing but the sober core spaces, 
and endowed with the Lawson topology, they are the corresponding fan spaces. We generalize the characterization 
of continuous lattices as meet-continuous lattices with T$_2$ Lawson topology and extend 
the Fundamental Theorem of Compact Semilattices to non-complete structures.
Finally, we investigate cardinal invariants like density and weight of the involved objects.

\vspace{1ex}




\end{abstract}

\maketitle

\vspace{-6ex}

\begin{center}
\includegraphics[height=7.5cm]{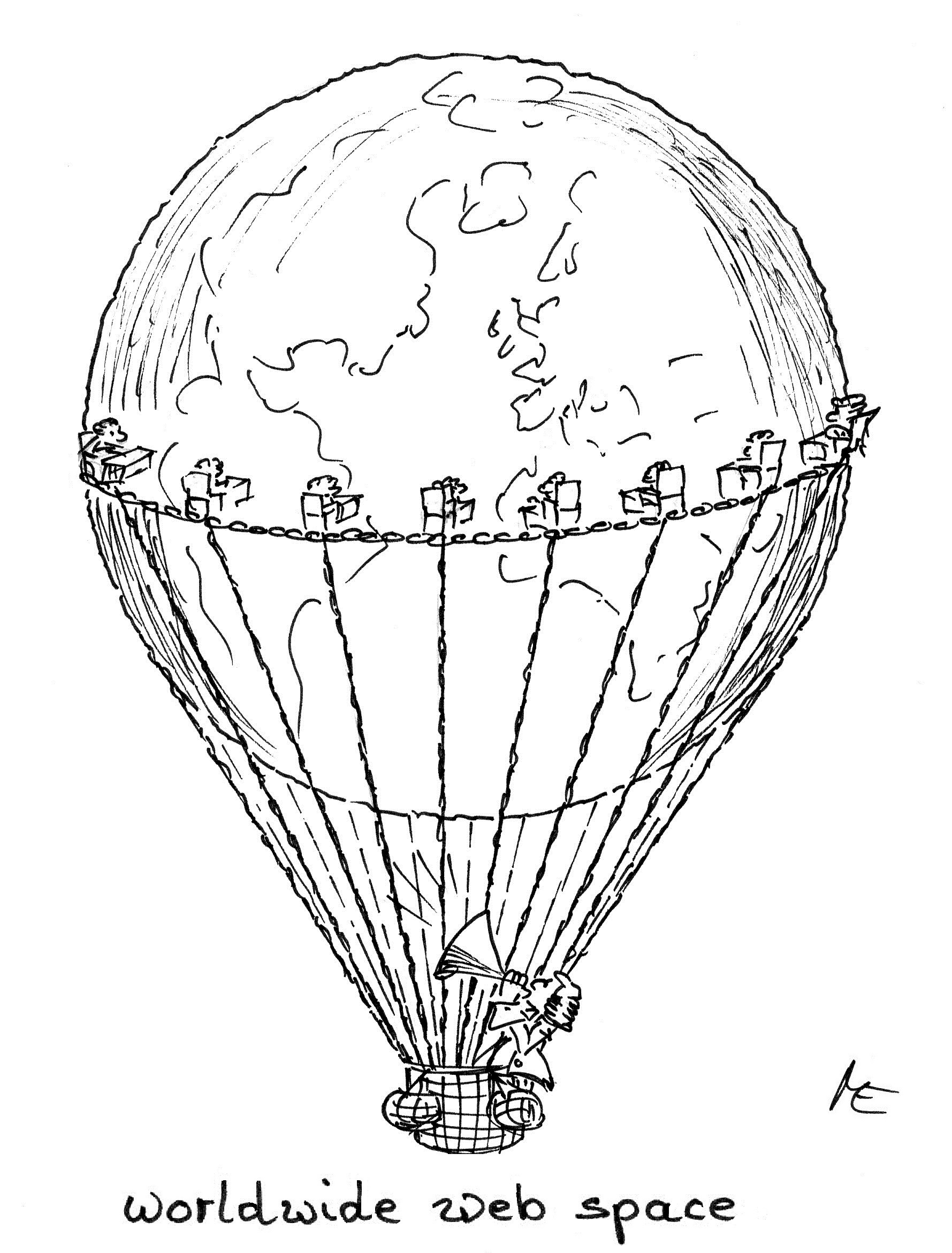}
\end{center}

\vspace{-4ex}

\newpage

\section{Introduction}
\label{intro}

Suitable models for the theory of computation and approximation are certain (quasi-)ordered sets, 
whose elements represent states of computation, knowledge or information, 
while the order abstractly describes refinement, improvement or temporal sequence.
Let us briefly record the relevant order-theoretical terminology.  

A {\em quasi-ordered set} or {\em qoset} is a pair $Q = (X,\leq)$ 
with a reflexive and transitive relation $\leq$ on $X$. The dual order is denoted by $\geq$, and the dual qoset $(X,\geq)$ by $\widetilde{Q}$.\\
If $\,\leq\,$ is antisymmetric, we speak of a {\em (partial) order} and an {\em ordered set} or a {\em poset}.\\
A {\em lower set}, {\em downset} or {\em decreasing set} is a subset $Y$ that coincides with its\,\,{\em down-closure}\,\,${\downarrow\!  Y}$, 
consisting of all $x \!\in\! X$ with $x \!\leq\! y$ for at least one $y \!\in\! Y$.
The {\em up-closure} ${\uparrow\! Y}$ and {\em upper sets (upsets, increasing sets)} are defined dually. 
The upper\,sets form the\,{\em upper Alexandroff topology}\,\,$\alpha Q$,
and the lower sets the {\em lower Alexandroff topology}\,\,$\alpha \widetilde{Q}$.
A set $D\subseteq Q$ is {\em (up-)directed}, resp.\ {\em filtered} or {\em down-directed}, 
if every finite subset of $D$ has an upper, resp.\ lower bound in\,$D$; in particular, $D$ cannot be empty.
An {\em ideal} of $D$ is a directed lower set, and a {\em filter} is a filtered upper set; for $x \!\in\! X$, 
the set ${\downarrow\!x} \!=\! {\downarrow\!\{ x\}}$ is the {\em principal ideal}, and ${\uparrow\!x} \!=\! {\uparrow\!\{ x\}}$ is
the {\em principal filter} generated by $x$.
A poset $P$ is called {\em up-complete}, {\em directed complete}, a {\em dcpo}, or a {\em cpo} 
if each directed subset $D$ or, equivalently, each ideal has a join, that is, a least upper bound ({\em supremum}), denoted by $\bigvee\! D$.
The {\em (ideal) up-completion} of a qoset $Q$ is $\I Q$, the set of all ideals, ordered by inclusion.
The {\em Scott topology} $\sigma P$ of a poset $P$ consists of all upper sets $U$ that meet any directed set having a join in\,\,$U$.

Both in the mathematical and in the computer-theoretically oriented literature (see e.g.\ \cite{AJ}, \cite{Edom}, \cite{CLD}, \cite{SLG}, \cite{V}), 
the word {\em `domain'} represents quite diverse structures, and in order-theoretical contexts, its meaning ranges 
from rather general notions like dcpos to quite specific kinds of posets like $\omega$-algebraic dcpos, sometimes with additional properties.
Here, we adopt the convention to call arbitrary up-complete posets {\em domains}
and to speak of a {\em continuous poset} if for each element $x$ there is a least ideal having a join above\,\,$x$.
Notice that our {\em continuous domains} are the {\em continuous posets} in \cite{Com} and \cite{Hoff2}, but they are the {\em domains} in \cite{CLD},
whereas our continuous posets and those in \cite{CLD} need not be up-complete. 

Although continuous domains usually are defined in order-theoretical terms, there exist also topological
descriptions of them, for example, as sober locally super\-compact spaces (Ern\'e \cite{EABC, Emin}, Hoffmann \cite{Hoff2}, Lawson \cite{CLD, Lss}). 
It is one of our main purposes in the subsequent investigations to drop the completeness or sobriety hypotheses 
without loosing relevant results applicable to domain theory.
The term {\em space} always means {\em topological space}, but extensions to arbitrary closure spaces are possible 
(see \cite{EABC} and \cite{Eclo}). 
Several classes of spaces may be characterized by certain infinite distribution laws for their lattices of open sets \cite{Eweb}. 
Recall that a {\em frame} or {\em locale} \cite{Jo} is a complete lattice $L$ satisfying the identity
\vspace{-.5ex}
$$
\textstyle{ {\rm (d)} \ x\wedge \bigvee Y = \bigvee \{ x\wedge y : y\in Y \} \vspace{-.5ex}}
$$
for all $x\in L$ and $Y \subseteq L$; the dual of (d) characterizes {\em coframes}. The identity
\vspace{-.5ex}
$$
\textstyle{{\rm (D)} \ \bigwedge \,\{ \bigvee Y : Y \!\in \Y\} = \bigvee\bigcap \Y} \vspace{-.5ex}
$$
for all collections $\Y$ of lower sets, defining {\em complete distributivity}, is much stronger.
However, frames may also be defined by the identity (D) for all {\em finite} collections $\Y$ of lower sets. 
An up-complete meet-semilattice satisfying (d) for all directed sets (or ideals) $Y$ is called {\em meet-continuous}. 
Similarly, the {\em continuous lattices} in the sense of Scott \cite{Com}, \cite{Sco} are the complete lattices enjoying the identity (D) for ideals instead of lower sets.
Therefore, completely distributive lattices are also called {\em super\-continuous};
alternative descriptions of complete distributivity by equations involving choice functions are equivalent to the Axiom of Choice\,\,(see\,\,\cite{Herr}).
A complete lattice satisfying (D) for all collections of finitely generated lower sets is called {\em $\F$-distributive} or a {\em wide coframe},
and its dual a {\em quasitopology} or a {\em wide frame} (cf.\ \cite{Eweb}).
A lattice is {\em spatial} iff it is isomorphic to a topology. All spatial lattices are wide frames.
For any space $(X,\S)$, the frame of open sets is $\S$, and the coframe of closed sets is denoted by $\S^c$.
The closure of a subset $Y$ is denoted 
by $cl_{\S}Y$ or $Y^-\!$, and the interior by $int_{\S}Y\!$ or $Y^{\circ}$. The {\em specialization order} is given by 
\vspace{-1ex}
$$
x\leq y \ \Leftrightarrow \ x\leq_{\S} y \ \Leftrightarrow \ x\in \{ y\}^- \ \Leftrightarrow \  \forall\,U\! \in \S\ (x\in U \,\Rightarrow \, y\in U).
$$  
It is antisymmetric iff $(X,\S)$ is T$_0$, but we speak of a specialization order also in the non--T$_0$ setting.
The {\em saturation} of a subset $Y$ is the intersection of all its neighborhoods, and this is the up-closure of $Y$ relative to the specialization order.
In the {\em specialization qoset}  $\Sigma^{-\!} (X,\S) = (X,\leq_{\S})$, the principal ideals are the point closures, and the principal filters are the cores, 
where the {\em core} of a point $x$ is the saturation of the singleton $\{ x\}$; the lower sets are the unions of cores, or of arbitrary closed sets, 
and the upper sets are the saturated sets. 
A topology $\S$ on $X$ is {\em compatible} with a quasi-order $\leq$ if $Q = (X,\leq)$ is the specialization qoset of $(X,\S)$ or, equivalently,
$\upsilon Q \subseteq \S  \subseteq \alpha Q$, where $\upsilon Q$ is the {\em weak upper topology}, generated by the complements of principal ideals;
the {\em weak lower topology} of $Q$ is the weak upper topology $\upsilon \widetilde{Q}$ of the order-dual $\widetilde{Q}$.
Of course, in other contexts, compatibility of a topology with an order relation may have a different meaning (cf.\ \cite[VI]{CLD}).

In \cite{Eweb}, we have introduced three classes of spaces that might be useful for the mathematical foundation of communication and information theory 
(order-theoretical notions refer to the specialization order):
\BIT
\IT[--] {\em web spaces} have neighborhood bases of webs at each point $x$, i.e.\ unions of filtered sets each of which contains $x$,
\IT[--] {\em wide web spaces} have neighborhood bases of filtered sets at each point,
\IT[--] {\em worldwide web spaces} or {\em core spaces} have neighborhood bases of principal filters (cores) at each point.
\EIT
As shown in \cite{Eweb}, each of these three classes of spaces may be described by an infinite distribution law for their topologies: a space is a
\vspace{-1ex}
\BIT
\IT[--] web space iff its topology is a coframe,
\IT[--] wide web space iff its topology is a wide coframe,
\IT[--] worldwide web space iff its topology is completely distributive.
\EIT

In Section \ref{convex}, we briefly review the construction of patch spaces and some applications to web spaces, as developed in \cite{Epatch}.
The patch spaces of a given space are obtained by joining its topology with a {\em cotopology} 
(in \cite{Lss}: \mbox{{\em complementary topology}),} that is, a topology having the dual specialization order.
Useful for patch constructions are so-called {\em coselections} $\zeta$, which choose for any topology $\S$ a subbase $\zeta \S$ of a cotopology $\tau_{\zeta}\S$.
The topology $\S^{\zeta}$ generated by $\S \cup \zeta \S$ is then a patch topology, 
and the corresponding (quasi-ordered!) {\em $\zeta$-patch space} is ${\rm P}_{\zeta}(X,\S) = (X, \leq_{\S} ,\S^{\zeta})$.
As demonstrated in \cite{Epatch}, web spaces may be characterized by the property that their open sets are exactly the up-closures of the open sets in any patch space.

For us, compactness does {\em not} include the Hausdorff separation axiom T$_2$. 
Locally compact spaces undoubtedly form one of the most important classes of topological spaces. 
In the non-Hausdorff setting, one has to require whole bases of compact neighborhoods at each point, 
because one compact neighborhood for each point would not be enough for an efficient theory.
In certain concrete cases, one observes that each point of the space 
under consideration has even a neighborhood base consisting of {\em supercompact} sets,
i.e.\ sets each open cover of which has already one member that contains them. 
Such spaces occur, sometimes unexpectedly, in diverse fields of mathematics\,--\,not only topological but also algebraic ones\,--\,and in theoretical computer science. 
Section \ref{CDS} is devoted to a closer look at such {\em locally supercompact spaces}; they are nothing but the core spaces, because the supercompact saturated sets are just the cores. 
These spaces have been introduced in \cite{EABC}, where the name {\em core spaces} referred to the larger class of closure spaces,
and discussed further in \cite{Emin} and \cite{Eweb}; core spaces are also called {\em C-spaces} (in \cite{KL}: {\em c-spaces}),
but that term has a different meaning in other contexts (e.g.\ in \cite{Kue} and \cite{Pri}).

The core spaces are exactly the locally compact wide web spaces, but also the locally hypercompact web spaces, 
where a set is {\em hyper\-compact} if its saturation is finitely generated; while the interior operator of a web space preserves {\em finite} unions of saturated sets, 
the interior operator of a core space preserves {\em arbitrary} unions of saturated sets \cite{Eweb}. 
Moreover, the category of core spaces has a strong order-theoretical feature, 
being concretely isomorphic to a category of generalized quasi-ordered sets (see \cite{EABC} and Section \ref{CDS} for precise definitions and results).  
Core spaces share useful properties with the more restricted {\em basic spaces} or {\em B-spaces} 
(having a least base, which then necessarily consists of all open cores) and with the still more limited {\em Alexandroff-discrete spaces} 
or {\em A-spaces} (in which all cores are open) \cite{Alex}, \cite{EABC}, \cite{Emin};
but, in contrast to A- and B-spaces, core spaces are general enough to cover important examples of classical analysis. For instance, the Euclidean topology on 
${\mathbb R}^n$ (which, ordered componentwise, is a continuous poset but not a domain!) 
is the weak(est) patch topology of the Scott topology, which makes ${\mathbb R}^n$ a core space. 

In Section \ref{sectorspaces}, we characterize the patch spaces of core spaces as {\em sector spaces}. 
These are {\em ${\uparrow}$-stable semi-qospaces} (meaning that up-closures of open sets are open, and principal ideals and principal filters are closed) 
with neighborhood bases of so-called {\em sectors}, 
a special kind of webs having least elements. The restriction of the patch functor ${\rm P}_{\zeta}$ to the category of core spaces yields a 
concrete isomorphism to the category of {\em $\zeta$-sector spaces}, which fulfil strong convexity and separation axioms. 
In particular, the {\em weak patch functor} ${\rm P}_{\upsilon}$ induces a concrete \mbox{categorical} isomorphism between core spaces and {\em fan spaces},
i.e.\ ${\uparrow}$-stable semi-qospaces in which each point has a neighborhood base of 
{\em fans} ${\uparrow\!u}\setminus{\uparrow\!F}$ with finite sets\,\,$F$. 
In Section \ref{fanspaces}, such fan spaces are investigated and characterized by diverse order-topological properties. 

Our considerations have useful consequences for topological aspects of domain theory, as the continuous domains, equipped with the Scott topology, 
are nothing but the sober core spaces, and these correspond to fan spaces that carry the {\em Lawson topology}, the weak patch topology of the Scott topology.
We find alternative descriptions of such ordered spaces, including convexity properties, separation axioms and conditions on the interior operator. 
This enables us to generalize the characterization of continuous lattices as meet-continuous lattices whose Lawson topology is Hausdorff \cite[III--2.11]{CLD} and
the Fundamental Theorem of Compact Semilattices \cite[VI--3]{CLD} to non-complete situations.
Crucial is the fact that a semilattice with a compatible topology is semitopological iff it is a web space, 
and (locally compact) topological with small semilattices iff it is a (world) wide webspace.

The category of T$_0$ core spaces and that of ordered fan spaces are not only equivalent to the category C-ordered sets, but also to the category of {\em based domains}, i.e.\ pairs 
consisting of a continuous domain and a basis of it (in the sense of \cite{CLD}).

In the last section, we study weight and density of the spaces under consideration, 
using the order-theoretical description of core spaces \cite{EABC}. 
For example, the weight of a core space is equal to the density of any of its patch spaces, 
but also to the weight of the lattice of {\em closed} sets. This leads to the conclusion that
the weight of a completely distributive lattice is always equal to the weight of the dual lattice.

If not otherwise stated, all results are derived in a choice-free set-theoretical environment;
i.e., we work in {\sf ZF} or {\sf NBG} (Zermelo--Fraenkel or Neumann--Bernays--G\"odel set theory) but not in {\sf ZFC} (i.e.\ {\sf ZF} plus Axiom of Choice).
For basic categorical concepts, in particular, concrete categories, functors and isomorphisms, see Ad\'amek, Herrlich and Strecker \cite{AHS}.
For relevant order-theoretical and topological definitions and facts, refer to the monograph {\it Continuous Lattices and Domains} by 
G.\ Gierz, K.\,H.\ Hofmann, K.\ Keimel, J.\,D.\ Lawson, M.\ Mislove, and D.\,S.\ Scott\,\,\cite{CLD}.

\newpage

\section{Patch spaces and web spaces}
\label{convex}

A {\em (quasi-)\-ordered space} is a (quasi-)\-ordered set equipped with a topology. In this elementary definition,  
no separation properties and no relationship between order and topology are required. However, some classical separation axioms extend to the ordered case as follows.
A quasi-ordered space is a {\em lower semi-qospace} if all principal ideals are closed, 
an {\em upper qospace} if all principal filters are closed, and a {\em semi-qospace} if both conditions hold 
(these conditions mean that the quasi\-order is {\em lower semiclosed}, {\em upper semiclosed} or {\em semiclosed}, respectively, in the sense of \cite[VI-1]{CLD}). 
An ordered semi-qospace is a {\em semi-pospace} or {\em T$_1$-ordered\,}. 
A space equipped with a closed quasi-order $\leq$ (regarded as a subset of the square of the space) is called a {\em qospace}, 
and a {\em pospace} in case $\leq$ is a (partial) order \cite{CLD}. 
Alternatively, qospaces may be characterized by the condition that for $x\not\leq y$, there are open $U$ and $V$ with 
$x \!\in\! U$, $y \!\in\! V$, and ${\uparrow\!U} \mathop{\cap} {\downarrow\! V} = \emptyset$. 
Similarly, we define {\em T$_2$-ordered spaces} to be ordered spaces in which for $x \not\leq y$ 
there are an open upper set containing $x$ and a disjoint open lower set containing $y$;
some authors call such spaces {\em strongly $T_2$-ordered} and mean by a {\em T$_2$-ordered space} a pospace (cf.\ K\"unzi\,\,\cite{Kue}, McCartan\,\,\cite{McC}).
A quasi-ordered space is said to be {\em upper regular} if for each open upper set $O$ containing a point $x$,
there is an open upper set $U$ and an closed upper set $B$ such that $x\in U \subseteq B \subseteq O$, 
or equivalently, for each closed lower set $A$ and each $x$ not in $A$, 
there is an open upper set $U$ and a disjoint open lower set $V$ with $x\in U$ and $A\subseteq V$. 
An upper regular T$_1$-ordered space is said to be {\em upper T$_3$-ordered}. 
{\em Lower regular} spaces are defined dually. Note the following irreversible implications:

{\em compact qospace $\Rightarrow$ upper regular semi-qospace $\Rightarrow$ qospace $\Rightarrow$ semi-qospace,}\\
\indent {\em compact pospace $\,\Rightarrow$ upper\,T$_3$-ordered $\Rightarrow$\,T$_2$-ordered $\Rightarrow$ pospace $\Rightarrow$\,T$_1$-ordered$\,+T_2$.}

\vspace{.5ex}

\noindent For any quasi-ordered space $T = (Q,\T) = (X,\leq,\T)$,

\vspace{.5ex}

$\T^{\,\leq} = \T \cap \alpha Q\ $ is the topology of all open upper sets (also denoted by $\T^{\,\sharp}$),\\
\indent $\T^{\,\geq} = \T \cap \alpha \widetilde{Q}\ $ is the topology of all open lower sets \hspace{.6ex}(also denoted by $\T^{\,\flat}$).

We call ${\rm U} T = (X,\T^{\leq})$ the {\em upper space} and ${\rm L} T = (X, \T^{\geq})$ the {\em lower space} of\,\,$T$.
A basic observation is that for lower semi-qospaces, the specialization order of $\T^{\leq}$ is $\leq$, 
while for upper semi-qospaces, the specialization order of $\T^{\geq}$ is $\geq$.

Recall that a subset $Y$ of a qoset $Q$ is {\em (order) convex} iff it is the intersection of an upper and a lower set. 
A quasi-ordered space is {\em locally convex} if the convex open subsets form a base,
{\em strongly convex} if its topology $\T$ is generated by $\T^{\leq} \cup \T^{\geq}$,
and {\em $\zeta$-convex} if $\T$ is generated by $\T^{\leq} \mathop{\cup} \zeta (\T^{\leq})$, where $\zeta$ is a coselection (see the introduction). 
Specifically, $\upsilon$-convex quasi-ordered spaces are called {\em hyperconvex}.
Thus, hyperconvexity means that the sets $U\setminus {\uparrow\!F}$ with $U\in \T^{\leq}$ and $F$ finite form a base.
Observe that $\zeta$-convexity implies strong convexity, which in turn implies local convexity, but not conversely
(counterexamples are given in\,\,\cite{Epatch}).

Let $\zeta$ be any coselection. A space $(X,\S)$ is said to be {\em $\zeta$-determined} if $\,\S^{\,\zeta\leq} = \S$.
A map between spaces is called {\em $\zeta$-proper} if it is continuous and preimages of closed sets 
relative to the $\zeta$-cotopology are $\zeta$-patch closed (whence such a map is $\zeta$-patch continuous); 
and a map between quasi-ordered spaces is {\em lower semicontinuous} if preimages of closed lower sets are closed. 
In \cite{Epatch}, many examples and counterexamples concerning these notions are discussed, and the following facts are established:

\vspace{-.5ex}

\BL
\label{pat}
The strongly convex semi-qospaces are exactly the patch spaces (of their upper spaces). 
The patch functor ${\rm P}_{\zeta}$ associated with a coselection $\zeta$ induces a concrete functorial isomorphism 
between the category of $\zeta$-determined spaces with continuous (resp.\ $\zeta$-proper) maps and that
of $\zeta$-convex semi-qospaces with isotone lower semicontinuous (resp.\ continuous) maps;
the inverse isomorphism is induced by the concrete upper space functor ${\rm U}$, sending a semi-qospace $(X,\leq, \T)$ to $(X,\T^{\leq})$. 
\EL

\vspace{-.5ex}

Any {\em upset selection} $\zeta$, assigning to each qoset $Q$ a collection $\zeta Q$ of upper sets such that 
${\uparrow\!x} = \bigcap\,\{ V\in \zeta Q : x\in V\}$ for all $x$ in $Q$,
gives rise to a coselection by putting $\zeta \S = \zeta \widetilde {Q}$ for any space $(X,\S )$ with specialization qoset $Q$. If each $\zeta Q$ is a topology,
we call $\zeta$ a {\em topological (upset) selection}; the largest one is $\alpha$, while the smallest one is\,\,$\upsilon$. 
By Lemma \ref{pat}, the {\em weak patch functor} ${\rm P}_{\upsilon}$ induces an isomorphism between 
the category of $\upsilon$-determined spaces and that of hyperconvex semi-qospaces.

An important intermediate topological selection is $\sigma$, where $\sigma Q$ is the {\em Scott topology}, consisting of all upper sets $U$ 
that meet every directed subset having a least upper bound that belongs to\,\,$U$ (in arbitrary qosets, 
$y$ is a least upper bound of $D$ iff $D\subseteq {\downarrow\! z} \Leftrightarrow y\leq z$).
The weak patch topology of $\sigma Q$ is the {\em Lawson topology} $\lambda Q = \sigma Q ^{\,\upsilon}$.
We denote by $\Sigma Q$ the {\em Scott space} $(X,\sigma Q )$ and by $\Lambda Q$ the (quasi-ordered) {\em Lawson space} $(Q,\lambda Q )$, 
whose upper space in turn is $\Sigma Q$. Thus, all Scott spaces are $\upsilon$-determined, and all Lawson spaces are hyperconvex semi-qospaces.

A quasi-ordered space $(Q,\T)$ is said to be {\em upwards stable} or {\em ${\uparrow}$-stable}
if it satisfies the following equivalent conditions:  
\vspace{-.5ex}
\BIT
\IT[{\rm (u1)}] $O\in \T$ implies ${\uparrow\!O}\in \T$.
\IT[{\rm (u2)}] $\T^{\leq} = \{ {\uparrow\! O} : O\in \T\}$.
\IT[{\rm (u3)}] The interior of each upper set is an upper set: $int_{\T} Y\! = int_{\T^{\leq}}Y$ if $Y \!= {\uparrow\!Y}$.
\IT[{\rm (u4)}] The closure of each lower set is a lower set: $cl_{\T} Y\! = cl_{\T^{\leq}}Y$ if $Y \!= {\downarrow\!Y}$.
\EIT

A {\em web} around a point $x$ in a qoset is a subset containing $x$ and with each point $y$ a common lower bound of $x$ and $y$;
if $\leq$ is a specialization order, that condition means that the closures of $x$ and $y$ have a common point in the web. 
By a {\em web-(quasi-)ordered space} we mean an ${\uparrow}$-stable (quasi-)ordered space in which every point has a neighborhood base of webs around it.
In the case of a space equipped with its specialization order, this is simply the definition of a {\em web space}.
Many characteristic properties of web spaces and of web-quasi-ordered spaces are given in \cite{Eweb} and \cite{Epatch}. 
Note that the {\em meet-continuous dcpos} in the sense of \cite{CLD} are just those domains whose Scott space is a web space \cite{Eweb}.
The following result from \cite{Epatch} underscores the relevance of web spaces for patch constructions:

\BPR
\label{webpatch}
A space $S$ is a web space iff all its patch spaces are web-quasi-or\-dered and have the upper\,space\,\,$S$.
Conversely, the strongly convex web-quasi-ordered semi-qospaces are the patch spaces of their upper spaces, and these are web spaces.
For any coselection $\zeta$, the patch functor ${\rm P}_{\zeta}$ induces a concrete isomorphism between the category of 
web spaces and that of $\zeta$-convex, web-quasi-ordered semi-qospaces.
\EPR

We now are going to derive an analogous result for wide web spaces; the case of worldwide web spaces (core spaces) is deferred to the next section.
The notion of wide web spaces is a bit subtle: while in a web space every point has a neighborhood base consisting of {\em open} webs around it \cite{Eweb}, 
in a wide web space it need not be the case that any point has a neighborhood base consisting of {\em open} filtered sets; 
the spaces with the latter property are those which have a `dual' (Hoffmann \cite{Hoff1}), that is, whose topology is dually isomorphic to another topology;
see \cite{Eweb} for an investigation of such spaces and a separating counterexample.  Note the implications

\vspace{.5ex}

{\em completely distributive $\,\Rightarrow\,\ $ dually spatial $\,\ \ \Rightarrow\ $ wide coframe $\hspace{1.8ex}\Rightarrow\, $ coframe}

\vspace{.5ex}

\noindent and the corresponding (irreversible) implications for spaces:

\vspace{.5ex}

{\em worldwide web space $\ \ \ \Rightarrow\ $ space with dual $\,\Rightarrow\, $ wide web space $\,\Rightarrow\, $ web space.}

\vspace{.5ex}

\noindent It is now obvious to introduce an ordered version of wide web spaces by calling a 
quasi-ordered space {\em locally filtered} if each point has a base of filtered neighborhoods.

\BT
\label{wideweb}
A space $S$ is a wide web space iff all its patch\,spaces\,are\,locally fil\-tered and\,up\-stable 
with\,upper\,space\,$S$.\,The strongly\,convex, locally filtered\,\,${\uparrow}$-stable semi\-qospaces 
are the patch spaces of their upper spaces, and these are are wide\,web\,spaces. 
Each patch functor ${\rm P}_{\zeta}$ induces a concrete isomorphism between the category of wide web spaces 
and that of $\zeta$-convex, locally filtered, ${\uparrow}$-stable semi-qospaces. 
\ET

\BP
Let $S \!=\! (X,\S )$ be a wide web space and $T \!=\! (X,\leq, \T )$ a patch space of\,\,$S$. 
By Proposition \ref{webpatch}, $T$ is ${\uparrow}$-stable, and $S$ is the upper space of $T$, i.e., $\S = \T^{\leq}$.
Given $x \!\in\! O \!\in\! \T$, find $U \!\in\! \S$ and $V \!\in\! \T^{\geq}$ with $x \in U \cap V \subseteq O$, 
and a filtered set $D$ with $x \in W = int_{\S} D \subseteq D \subseteq U$.
Then, $D\cap V$ is filtered (since $V = {\downarrow\!V}$) 
with $x\in W\cap V \subseteq D\cap V \subseteq U \cap V \subseteq O$, and $W\cap V \in \T$. Thus, $T$ is locally filtered. 

Conversely, let $T = (X, \leq, \T )$ be a strongly convex, locally filtered and ${\uparrow}$-stable semi-qospace.
Then $T$ is web-quasi-ordered and, by Proposition \ref{webpatch}, a patch space of the web space $(X,\T^{\leq})$.
For $x \in O \in \T^{\leq}$, there is a filtered $D\subseteq O$ with $x \in int_{\T}D$.
Then ${\uparrow\!D}$ is filtered, $x \in W = {\uparrow\! int_{\T} D} \subseteq {\uparrow\!D} \!\subseteq\! O$,
and by ${\uparrow}$-stability, $W \in \T^{\leq}$; thus, $(X,\T^{\leq})$ is a wide web space. The rest follows from Proposition \ref{pat}.
\EP

\BCO
\label{webco}
The strongly convex, locally filtered and ${\uparrow}$-stable T$_1$-ordered spaces are exactly the patch spaces of T$_0$ wide web spaces.
\ECO

\section{Core spaces and C-quasi-ordered sets}
\label{CDS}

Strengthening the notion of compactness, we call a subset $C$ of a space {\em super\-compact} 
if every open cover of $C$ has a member that contains $C$; and {\em local supercompactness} 
means the existence of supercompact neighborhood bases at each point. As mentioned in the introduction, an equivalent condition is 
that each point has a neighborhood base of cores (possibly of different points!)\,--\,in other words, that we have a {\em core space}. 
Under the assumption of the Ultrafilter Theorem (a consequence of the Axiom of Choice), 
many properties of locally super\-compact spaces are shared by the more general {\em locally hypercompact spaces},
where a subset is called {\em hypercompact} if its saturation is generated by a finite subset. 
In view of the next proposition, proven choice-freely in \cite{EABC} and \cite{Eweb}, core spaces may be viewed as an infinitary analogue 
of web spaces (whence the name {\em `worldwide web spaces'}\,). 

\BPR
\label{supertop}
For a  space $S = (X,\S)$, the following conditions are equivalent:
\vspace{-.5ex}
\BIT
\IT[{\rm (1)}] $S$ is a core space.
\IT[{\rm (2)}] $S$ is locally supercompact.
\IT[{\rm (3)}] The lattice of open sets is supercontinuous (completely distributive).
\IT[{\rm (4)}] The lattice of closed sets is supercontinuous.
\IT[{\rm (5)}] The lattice of closed sets is continuous.
\IT[{\rm (6)}] The interior operator preserves arbitrary unions of upper sets.
\IT[{\rm (7)}] The closure operator preserves arbitrary intersections of lower sets.
\IT[{\rm (8)}] $S$ is a locally hypercompact web space.
\IT[{\rm (9)}] $S$ is a locally compact wide web space.
\EIT
\EPR

Core\,spaces have pleasant properties. For example, on account of (6) resp.\,(7), the interior resp.\ closure operator of a core space induces a complete homomorphism 
from the completely distributive lattice of upper resp.\,lower sets onto the lattice of open resp.\,closed sets.
In \cite{BrE}, it is shown that a nonempty product of spaces is a core space iff all factors are core spaces and all but a finite number of them are supercompact.
Similarly, a nonempty product of spaces is a (wide) web space iff all factors are (wide) web spaces and all but a finite number are filtered. 
Assuming the Principle of Dependent Choices (another consequence of the Axiom of Choice), 
one can show that all core spaces have a dual (a base of filtered open sets; see \cite{Eweb}).

Computationally convenient is the fact that core spaces are in bijective correspondence to so-called {\em idempotent ideal relations} or {\em C-quasi-orders}.
These are not really quasi-orders but idempotent relations $\Ro$ (satisfying $x\,\Ro\,z \Leftrightarrow \exists\, y \,(x\,\Ro\,y\,\Ro\,z)$) on a set $X$ so that the sets
\vspace{-1ex}
$$\Ro y = \{ x \in X : x \,\Ro\, y\} \ \ (y\in X)
$$ 
are ideals with respect to the {\em lower quasi-order} $\leq_{\Ro}$ defined by
\vspace{-.5ex}
$$
x\leq_{\Ro} y \ \Leftrightarrow \Ro x \subseteq \Ro y.
$$
And $\Ro$ is called a {\em C-order} if, moreover, $\leq_{\Ro}$ is an order, i.e., $\Ro y \!=\! \Ro z$ implies $y \!=\! z$. 
The pair $(X,\Ro )$ is referred to as a {\em C-(quasi-)ordered set} \cite{EABC}. 
Any C-order $\Ro$ is an {\em approximating auxiliary relation} for the poset $(X,\leq_{\Ro})$ in the sense of \cite{CLD}.

For an arbitrary relation $\Ro$ on a set $X$ and any subset $Y$ of $X$, put
$$\Ro\, Y = \{ x\in X : \exists\,y\in Y\, (x\,\Ro\,y)\} , \  Y\Ro = \{ x\in X : \exists\,y\in Y\, (y\,\Ro\,x)\},\vspace{-.5ex}
$$
$$\ _{\Ro}\O = \{ \,\Ro\,Y : Y\subseteq X\} ,  \ \O_{\Ro} = \{ Y\Ro : Y\subseteq X\}. \  
$$
$_{\Ro}\O $ and  $\O_{\Ro}$ are closed under arbitrary unions,
and in case $\Ro$ is an ideal relation, $\O_{\Ro}$ is even a topology.
On the other hand, if $\Ro$ is idempotent then $_{\Ro}\O$ consists of all rounded subsets, 
where a subset $Y$ is said to be {\em round(ed)}  if $Y = \Ro Y$ \cite{GK}, \cite{Lri}.

Typical examples of C-orders are the {\em way-below relations} $\ll$ of {\em continuous posets} $P$, 
in which for any element $y$ the set 
\vspace{-.5ex}
$$\textstyle{\ll\! y = \{ x\in P : x \ll y\} = \bigcap \,\{ D\in \I P : \bigvee\! D \mbox{ exists and } y\leq \bigvee\! D\}}
$$
is an ideal (called the {\em way-below ideal of} $y$) with join $y$ (recall that $\I P$ is the set of all ideals). 
The next result, borrowed  from \cite{EHab}, leans on the {\em interpolation property} of continuous posets, saying that their way-below relation is idempotent.

\BL
\label{conti}
For every continuous poset, the way-below relation is a C-order $\Ro$ such that for directed subsets $D$ (relative to $\leq_{\Ro}$), 
$y = \bigvee\! D$ is equivalent to $\Ro y = \Ro D$; and any C-order with that property is the way-below relation of a continuous poset. 
\EL

\BX
\label{Exnotcon}
{\rm
The following sublattice $L$ of the plane $\mathbb{R}^2$ has a unique compatible topology $\alpha L = \upsilon L$, and the interval topology $\iota L$ is discrete \cite{Epatch}.

\begin{picture}(250,150)

\put(-10,120){$L = \{ a_n,\,b_n,\,c_n: n\in \omega\}$}

\put(-10,100){with}

\put(-10,80){$a_n = (0,-2^{-n}),$}

\put(-10,60){$b_n =(2^{-n},0),$}

\put(-10,40){$c_n = (2^{-n},2^{-n}).$}

\put(110,0)
{
\begin{picture}(200,150)
\put(70,42){$L$}
\put(15,10){$a_0$}
\put(78,80){$b_0$}
\put(78,130){$c_0$}

\put(74,77){\line(0,1){59}}
\put(10,10){\circle{4}}
\put(10,12){\line(0,1){29}}
\put(10,43){\circle{4}}
\put(10,45){\line(0,1){13}}
\put(10,60){\circle{4}}
\put(10,66){\circle{1}}
\put(10,69){\circle{1}}
\put(10,72){\circle{1}}
\put(10,75){\circle{1}}
\put(13,75){\circle{1}}
\put(16,75){\circle{1}}
\put(19,75){\circle{1}}
\put(13,78){\circle{1}}
\put(16,81){\circle{1}}
\put(19,84){\circle{1}}

\put(24,75){\circle{4}}
\put(26,75){\line(1,0){13}}
\put(24,77){\line(0,1){9}}
\put(41,75){\circle{4}}
\put(41,77){\line(0,1){26}}
\put(43,75){\line(1,0){29}}

\put(24,88){\circle{4}}
\put(26,90){\line(1,1){13}}
\put(41,105){\circle{4}}
\put(43,107){\line(1,1){29}}

\put(74,75){\circle{4}}
\put(74,138){\circle{4}}

\end{picture}
}

\put(240,0)
{
\begin{picture}(100,147)
\put(70,42){$\widetilde{L}$}
\put(55,135){$a_0$}
\put(-8,71){$b_0$}
\put(-8,12){$c_0$}

\put(6,73){\line(0,-1){59}}
\put(70,140){\circle{4}}
\put(70,138){\line(0,-1){29}}
\put(70,107){\circle{4}}
\put(70,105){\line(0,-1){13}}
\put(70,90){\circle{4}}
\put(70,84){\circle{1}}
\put(70,81){\circle{1}}
\put(70,78){\circle{1}}
\put(70,75){\circle{1}}
\put(67,75){\circle{1}}
\put(64,75){\circle{1}}
\put(61,75){\circle{1}}
\put(67,72){\circle{1}}
\put(64,69){\circle{1}}
\put(61,66){\circle{1}}

\put(56,75){\circle{4}}
\put(54,75){\line(-1,0){13}}
\put(56,73){\line(0,-1){9}}
\put(39,75){\circle{4}}
\put(39,73){\line(0,-1){26}}
\put(37,75){\line(-1,0){29}}

\put(56,62){\circle{4}}
\put(54,60){\line(-1,-1){13}}
\put(39,45){\circle{4}}
\put(37,43){\line(-1,-1){29}}

\put(6,75){\circle{4}}
\put(6,12){\circle{4}}
\end{picture}
}
\end{picture}

\vspace{-1.5ex}

\noindent $L$ and its dual $\widetilde{L}$ are trivially continuous, having no non-principal ideals possessing a join, so that 
the way-below relation is the order relation in these lattices. 
But, being incomplete, they are not continuous lattices in the sense of Scott \cite{Com}, \cite{Sco}.
The completion of $L$ by one `middle' point $(0,0)$ is a continuous but not super\-continuous frame, while the completion of $\widetilde{L}$ is not even meet-continuous. 
}
\EX


The aforementioned one-to-one correspondence between core spaces and C-quasi\-orders is based on the following remark and definition: 
every space $(X,\S)$ carries a transitive (but only for A-spaces reflexive) {\em interior relation} $\Ro_{\S}$, given by
\vspace{-.5ex}
$$
x \,\Ro_{\S}\,y \, \Leftrightarrow \, y \in ({\uparrow\!x})^{\circ} = int_{\S} ({\uparrow\!x}), \vspace{-.5ex}
$$
where ${\uparrow\!x} = \bigcap\, \{ U \in \S : x \in U\}$ is the core of $x$. Note that for any subset $A$ of $X$,\\
$\Ro_{\S}A$ is a lower set in $(X, \leq_{\S})$,
as $\,x \!\leq\! y\,\Ro_{\S} \, z$ entails  $z\in ({\uparrow\!y})^{\circ} \!\subseteq\! ({\uparrow\!x})^{\circ}$, hence $x \, \Ro_{\S} \,z$.\\
A map $f$ between C-quasi-ordered sets $(X,\Ro)$ and $(X',\Ro')$ {\em interpolates} if
for all $y \!\in\! X$ and $x' \!\in\!  X'$ with $x' \Ro\, f(y)$ there is an $x \in X$ with $x' \Ro' f(x)$ and $x \,\Ro\, y$, 
and $f$ is {\em isotone} (preserve the lower quasi-orders) iff $x\leq_{\Ro} y$ implies $f(x) \leq_{\Ro '}\! f(y)$.

\BT
\label{Cspace}
{\rm (1)} $(X,\S)$ is a core space iff there is a C-quasi-order $\Ro$ with $\S = \O_{\Ro}$.
In that case, $\Ro$ is the interior relation $\Ro_{\S}$, and $\leq_{\Ro}\!$ is the specialization order $\leq_{\S}$.

{\rm (2)} By passing from $(X,\S)$ to $(X,\Ro_{\S})$, and in the opposite direction from $(X,\Ro)$ to $(X,\O_{\Ro})$,
the category of core spaces with continuous maps is concretely isomorphic to the category of C-quasi-ordered sets with interpolating isotone maps.

{\rm (3)} For each closed set $A$ in a core space $(X,\S)$, the set $\Ro_{\S}A$ is the least lower set with closure $A$.
The closure operator induces an isomorphism between the super\-continuous lattice $_{\Ro}\O$ of all rounded sets and that of all closed sets, 
while $\O_{\Ro}$ is the supercontinuous lattice of all open sets. 

{\rm (4)} The irreducible closed subsets of a core space are exactly the closures of directed sets (in the specialization qoset). 
For an irreducible closed set\,$A$, the set $\Ro_{\S}A$ is the least ideal with closure $A$.\,The 
closure operator induces an iso\-morphism between the continuous domain of rounded ideals and that of irreducible closed sets.

{\rm (5)} The topology of a core space $(X,\S)$ is always finer than the Scott topology on its specialization qoset.

{\rm (6)} The cocompact topology of a core space $(X,\S)$ 
is the weak lower topology of the specialization qoset.
The patch topology $\S^{\pi}$ agrees with the weak patch topology\,\,$\S^{\upsilon}$.  
\ET

\BP
(1) and (2) have been established in \cite{EABC}. 
There are several reasonable alternative choices for the morphisms. For example, the {\em quasiopen} maps 
(for which the saturations of images of open sets are open) between core spaces 
are the relation preserving isotone maps between the associated C-quasi-ordered sets (cf.\ \cite{EABC}, \cite{HM}). 

(3) Let $\Ro$ be the interior relation and $\leq$ the specialization order $\leq_{\Ro}$.
We prove the identity $(\Ro A)^- = A^-$, using idempotency of $\Ro$ in the equivalence * below:

$y\in (\Ro A)^- \ \Leftrightarrow \ \forall\, U\in \O_{\Ro}\ (y\in U \Rightarrow U\cap \Ro A \neq \emptyset)$

$\Leftrightarrow  \ \forall \, x\in X\ (x \,\Ro \,y \,\Rightarrow\, x\Ro \cap \Ro A \neq \emptyset) \ \stackrel{*}{\Leftrightarrow} \ 
\forall \, x\in X\ (x \,\Ro \,y \Rightarrow x\Ro \cap A \neq \emptyset)$

$\Leftrightarrow \ \forall\, U\in \O_{\Ro}\ (y\in U \Rightarrow U\cap A \neq \emptyset) \ \Leftrightarrow \ y\in A^-.$

\noindent In particular, $(\Ro A)^- = A$ in case $A$ is closed. On the other hand, 
any rounded set $Y$ satisfies the identity $Y = \Ro Y = \Ro (Y^-)$: since each $x\Ro$ is an open set, we have 

$x\in \Ro Y \ \Leftrightarrow \ x\Ro \cap Y \neq \emptyset \ \Leftrightarrow \ x\Ro \cap Y^- \neq \emptyset \ \Leftrightarrow \ x\in \Ro (Y^-).$ 

\noindent And if $Y$ is any lower set with $Y^-\! = A$ then $\Ro A = \Ro (Y^-) = \Ro Y \subseteq {\downarrow\!Y} = Y.$

(4) Recall that a subset $A$ is irreducible iff it is nonempty and $A \subseteq B\cup C$ implies $A \subseteq B$ or $A \subseteq C$ for any closed sets $B,C$.
Directed subsets and their closures are irreducible. 
The rounded ideals of a C-quasi-ordered set (or a core space) form a domain $\I_{\Ro} = \{ I\in \I (X,\leq_{\Ro}): I = \Ro I\}$,
being closed under directed unions.  It is easy to see that $I \ll J$ holds in $\I_{\Ro}$ iff there is an $ x\in J$ with $I\subseteq {\downarrow\!x}$,
and that the join of the ideal $\ll \!J = {\downarrow_{\I_{\Ro}}} \{ \Ro x : x\in J\}$ is $J$. Thus, $\I_{\Ro}$ is continuous (cf.\ \cite{GK},\,\cite{Lss}).
For the isomorphism claim, observe that the coprime rounded sets are the rounded ideals, the coprime closed sets are the irreducible ones, 
and a lattice isomorphism preserves coprimes (an element $q$ is {\em coprime} iff the complement of ${\uparrow\!q}$ is an ideal).

(5) By (4), $I = \Ro_{\S} y$ is an ideal with ${\downarrow\! y} = I^-$, so $y$ is a least upper bound of $I$; now,
$y \in U \in \sigma (X,\leq)$ implies $U\cap \Ro_{\S} y \not = \emptyset$, i.e.\ $y\in U\Ro_{\S}$. Thus, $U = U\Ro_{\S}\in \S$.

(6) Let $C$ be a compact saturated set (i.e.\ $C \!\in\! \pi \S$) and $y \!\in\! X\setminus C$. 
There is an open neighborhood $U$ of $C$ not containing $y$. By local supercompactness\,of\,\,$(X,\S)$, we
have $U = \bigcup\, \{ int_{\S}({\uparrow\! x}) : x\in U\}$, and by compactness of $C$, we find a finite $F \subseteq U$ with
$C \!\subseteq \bigcup\, \{ int_{\S} ({\uparrow\! x}) : x \!\in\! F\} \!\subseteq\! {\uparrow\! F} \!\subseteq\! U$. 
For $Q = (X,\leq)$, ${\uparrow\! F}$ is $\upsilon\widetilde{Q}$-closed and does not contain $y$. 
Thus, $C$ is $\upsilon\widetilde{Q}$-closed, and $\pi \S$ is contained in $\upsilon \widetilde{Q}$; the reverse
inclusion is obvious, since cores (principal filters) are compact and saturated. 
\EP

Recall that a {\em sober} space is a T$_0$-space whose only irreducible closed sets are the point closures; 
and a {\em monotone convergence space} ({\em mc-space}) or {\em d-space} is a T$_0$ space 
in which the closure of any directed subset is the closure of a singleton (see \cite{Emin}, \cite{CLD}, \cite{Wy}). 
Now, from Theorem \ref{Cspace}, one deduces (cf.\ \cite{EABC}, \cite{Hoff2}, \cite{Law}, \cite{Lss}):

\BCO
\label{contdom}
The following conditions on a space $S$ are equivalent:
\vspace{-.5ex}
\BIT
\IT[{\rm (1)}] $S$ is a sober core space.
\IT[{\rm (2)}] $S$ is a locally supercompact monotone convergence space (d-space).
\IT[{\rm (3)}] $S$ is the Scott space of a (unique) continuous domain. 
\EIT  
\vspace{-.5ex}
The category of sober core spaces and the concretely isomorphic category of continuous domains are dual to the category of supercontinuous spatial frames.
\ECO

%
%

\vspace{-.5ex}

For the case of continuous posets that are not necessarily domains, see \cite{EScon}, \cite{Emin}.
Afficionados of domain theory might remark that continuous frames are automatically spatial (see \cite{CLD}, \cite{HL}). 
But that `automatism' requires choice principles, which we wanted to avoid in the present discussion. 
However, it seems to be open whether the spatiality of {\em supercontinuous} frames may be proved in a choice-free manner.

Since a T$_0$-space and its sobrification have isomorphic open set frames, it follows from Corollary \ref{contdom} that
a T$_0$-space is a core space iff its sobrification is the Scott space of a continuous domain. This completion process is reflected, via Theorem \ref{Cspace},
by the fact that the rounded ideal completion of a C-ordered set is a continuous domain, and the C-order 
is extended to its completion, meaning that $x\,\Ro\,y$ is equivalent to $\Ro x \ll \Ro y$ in the completion (cf. \cite{Emin}, \cite{CLD}, \cite{GK}, \cite{Lss}).

\section{Core stable spaces and sector spaces}
\label{sectorspaces}

As every core space is a web space, it is equal to the upper spaces of its patch spaces and therefore $\zeta$-determined for any coselection $\zeta$ (see Proposition \ref{webpatch}).
We now are going to determine explicitly these patch spaces, which turn out to have very good separation properties, 
whereas the only T$_1$ core spaces (in fact, the only T$_1$ web spaces) are the discrete ones. 
For alternative characterizations of such patch spaces, we need further properties of the interior operators. 
Call a quasi-ordered space $(Q, \T )$ {\em core stable} or {\em c-stable} if
$$
\textstyle{{\uparrow\!O} = \bigcup \,\{ \,int_{\T}({\uparrow\!u}) : u\in O\}  \ \mbox{ for each }O \in \T, }
$$
and {\em d-stable} if for any filtered (i.e.\ down-directed) subset $D$ of $Q$,
$$
\textstyle{int_{\T}D \subseteq \bigcup\,\{ int_{\T} \, ({\uparrow\!u}): u\in cl_{\T^{\geq}} D\}.}
$$
While core stability is a rather strong property, d-stability is a rather weak one (trivially fulfilled if all filters are principal). The terminology is justified by

\BL
\label{cstable}
Let $T = (X,\leq, \T)$ be a semi-qospace.

{\rm (1)} $T$ is d-stable whenever its lower space ${\rm L} T$ is a d-space.

{\rm (2)} $T$ is core stable iff $\,{\rm U}T = (X,\{ {\uparrow\!O} : O \in {\mathcal T}\})$ is a core space.

{\rm (3)} $T$ is core stable iff it is upper regular, locally filtered, ${\uparrow}$-stable and d-stable.   
\EL

\BP
(1) If the lower space $\,{\rm L}T = (X,\T^{\geq})$ is a d-space then for any filtered set $D$ in $(X,\leq)$ there is a $u$ 
with ${\uparrow\!u} = cl_{\T^{\geq}} D$ and $\,int_{\T} D \subseteq int_{\T} ({\uparrow\!u})$. 
Without proof, we note that ${\rm L}T$ is a d-space whenever $T$ is hyperconvex and $(X,\geq)$ is a domain. 

(2) Clearly, a core stable semi-qospace $(X,\leq ,\T)$ is ${\uparrow}$-stable, whence ${\mathcal T}^{\leq} = \{ {\uparrow\!O} : O \in {\mathcal T}\}$. 
For $U\in \T^{\leq}$ and $x\in U$ there exists a $u\in U$ with $x\in int_{\T} ({\uparrow\!u})$.
Then $W = {\uparrow int_{\T}({\uparrow\!u})}\in \T^{\leq}$ (by ${\uparrow}$-stability) and $x\in W\subseteq {\uparrow\!u} \subseteq {\uparrow\!U} = U$.
Since $\leq$ is the specialization order of $\T^{\leq}$, this ensures that $(X,\T^{\leq})$ is a core space.

Conversely, suppose $(X,\T^{\leq}) = (X, \{ {\uparrow\!O} : O \in {\mathcal T}\})$ is a core space. 
For $O\in {\mathcal T}$ and $y \in {\uparrow\!O}$, there is an $x\in {\uparrow\!O}$ and a $U\in \T^{\leq} \subseteq \T$ with $y\in U\subseteq {\uparrow\!x}$.
Now, pick a $u\in O$ with $x\in {\uparrow\!u}$; then $y\in U \subseteq {\uparrow\!x} \subseteq {\uparrow\!u}$, whence 
${\uparrow\!O} \subseteq \bigcup \,\{ \,int_{\T}({\uparrow\!u}) : u\in O\}$. 

(3) Let $T = (X,\leq, \T )$ be a core stable semi-qospace. For $x\in O\in \T^{\leq}$, there is a $u\in O$ with $x\in U = int_{\T} ({\uparrow\!u}) \subseteq {\uparrow\!u} \subseteq O$.
By ${\uparrow}$-stability, we have $U\in \T^{\leq}$, and since $T$ is a semi-qospace, ${\uparrow\!u}$ is a closed upper set. Hence, $T$ is upper regular. 

In order to check local filteredness, pick for $x\in O\in \T$ an element $u\in O$ with $x\in int_{\T} ({\uparrow\!u}) \mathop{\cap} O \subseteq {\uparrow\!u} \mathop{\cap} O\,$; 
this is a filtered set, possessing the least element\,\,$u$.

Moreover, $T$ is not only ${\uparrow}$-stable (see (2)) but also d-stable, since for any subset $D$ and any $x\in O = int_{\T} D$, 
there is a $u \in O \subseteq D \subseteq cl_{\T^{\geq}} D$ with $x\in int_{\T}({\uparrow\!u})$.

Conversely, suppose that $T$ is an upper regular, locally filtered, ${\uparrow}$-stable and d-stable semi-qospace. 
Then, for $O\in \T$ and $x\in {\uparrow\!O}$, we have:

$x\in {\uparrow\!O} \in \T^{\leq}$ \hfill (by ${\uparrow}$-stability),

there are $U\in \T^{\leq}$ and $B\in \T^{\geq c}$ with $x\in U \subseteq B \subseteq {\uparrow\!O}$ \hfill (by upper regularity),

a filtered $D$ with $x\in int_{\T} D \subseteq D \subseteq U$ \hfill (by local filteredness),

and an element $y\in cl_{\T^{\geq}} D$ with $x \in int_{\T} ({\uparrow\!y})$ \hfill (by d-stability).

\noindent It follows that $y \in cl_{\T^{\geq}} D \subseteq cl_{\T^{\geq}} U \subseteq B \subseteq {\uparrow\!O}$.
Now, choose a $u\in O$ with $u\leq y$, hence ${\uparrow\!y} \subseteq {\uparrow\!u}$, to obtain $x\in int_{\T} ({\uparrow\!u})$. 
Thus, ${\uparrow\!O} \subseteq \bigcup\,\{ int_{\T} ({\uparrow\!u}) : u\in O\}$, showing that $T$ is core stable.
\EP

By Lemma \ref{cstable}, every core stable semi-qospace is a qospace (being upper regular); in fact, core stability of a semi-qospace splits into the four properties

(c1) {\em upper regular} \ \ (c2) {\em locally filtered} \ \ (c3) {\em ${\uparrow}$-stable} \ \ (c4) {\em d-stable}.

\noindent These properties are independent: none of them follows from the other three.

\BX
\label{Ex31}
{\rm
Let $L$ be a non-continuous wide frame (for example, a T$_2$ topology that is not locally compact, like the Euclidean topology on ${\mathbb Q}$). 
As shown in \cite{Eweak}, $(L, \upsilon L)$ is a wide web space whose meet operation is continuous in the weak upper topology $\upsilon L$ and in the weak lower topology $\upsilon \widetilde{L}$, 
hence also in the interval topology $\iota L = \upsilon L\,^{\upsilon}$; and ${\rm I}L = (L,\iota L )$ is a topological meet-semilattice with small subsemilattices (see Section \ref{semi}).
Thus, it is locally filtered, ${\uparrow}$-stable and d-stable, since the lower space $(\widetilde{L},\upsilon \widetilde{L})$ is a d-space 
(for any complete lattice $L$). But the semi-pospace ${\rm I}L$ can be upper regular only if it is a pospace, hence T$_2$, which happens only if $L$ is a continuous lattice \cite{CLD}. 
Thus, ${\rm I}L$ satisfies (c2), (c3) and (c4), but not (c1). 

}
\EX

\BX
\label{Ex32}
{\rm 
For $0 < s < 1$, consider the non-compact subspace $S = [\,0, s\,[ \, \mathop{\cup}\, \{ 1\}$ of the Euclidean space ${\mathbb R}$, ordered by
$x\sqsubseteq y \Leftrightarrow x = y$ or $x = 0$ or $y=1$. As $\{ 1\}$ is clopen, it is easy to check that $S$ satisfies (c1), (c3) and (c4) (all filters are principal), 
but not (c2): no point except $1$ has a filtered neighborhood not containing 0. 
}
\EX

\BX
\label{Ex33}
{\rm
In Example 4.1 of \cite{Epatch} it is shown that the set

\begin{picture}(300,55)

\put(10,40){$X = \{ a,\top\} \cup \{ b_n : n\in \omega\}$, ordered by}

\put(10,20){$x\leq y \ \Leftrightarrow   x = y \mbox{ or } x = b_0 \mbox{ or } y = \!\top $}

\put(81,0){or $x = b_i, y = b_j, i < j,$}

\put(260,30){\circle{5}}
\put(250,28){$a$}
\put(262,32){\line(3,2){40}}
\put(262,28){\line(3,-2){40}}

\put(305,36){\circle{5}}
\put(305,20){\circle{5}}
\put(305,33){\line(0,-1){10}}
\put(305,18){\line(0,-1){15}}

\put(305,39){\line(0,1){5}}
\put(305,47){\circle{1}}
\put(305,51){\circle{1}}
\put(305,55){\circle{1}}

\put(305,60){\circle{5}}
\put(310,58){$\top$}
\put(305,0){\circle{5}}
\put(310,31){$b_2$}
\put(310,14){$b_1$}
\put(310,-4){$b_0$}

\end{picture}

\vspace{1ex}

\noindent is a complete but not meet-continuous lattice and a compact pospace when equipped with the Lawson topology. 
It satisfies (c1), (c2) and (c4) (because all filters are principal).
However, (c3) is violated, since $\{ a \}$ is open, while ${\uparrow\!a} = \{ a, \top\}$ is not. 
}
\EX

\BX
\label{Ex34}
{\rm
For infinite $I$, the function space ${\mathbb R}^I$ with the coordinatewise order and the Lawson topology satisfies (c1), (c2) and (c3), 
but its upper space $\Sigma ({\mathbb R}^I)$ is not a core space: 
otherwise, in view of the fact that bounded directed subsets converge to their suprema, 
${\mathbb R}^I$ would have to be a continuous poset (see \cite{Emin}), which holds only for finite $I$ (see \cite{EScon}).
Hence, by Lemma \ref{cstable}, (c4) must be violated.
}
\EX

Given a quasi-ordered space $(X,\leq,\T)$, we call any nonempty set of the form ${\uparrow\!u}\mathop{\cap} V$ with $V \!\in \!\T^{\,\geq}\!$  a {\em sector},
and a {\em $\zeta$-sector} if $V\in \zeta (\T^{\leq})$ for a coselection $\zeta$.
Hence, $u$ is the least element of the sector, and every sector is obviously a web around any point it contains.
By a ($\zeta$-){\em sector space} we mean an ${\uparrow}$-stable semi-qospace in which
each point $x$ has a base of ($\zeta$-)sector neighborhoods ${\uparrow\!u}\cap V$ (but the point $x$ need not be the minimum $u$ of such a sector).

\BT
\label{sectors}

The sector spaces are exactly the
\vspace{-.5ex}
\BIT
\IT[{\rm (1)}] patch spaces of core spaces,
\IT[{\rm (2)}] strongly convex, core stable (semi-)qospaces,
\IT[{\rm (3)}] strongly convex, upper regular, locally filtered, ${\uparrow}$-stable and d-stable qospaces. 
\EIT
In particular, all ordered sector spaces are T$_3$-ordered, a fortiori T$_2$-ordered.

\noindent Specifically, for any coselection $\zeta$, the $\zeta$-sector spaces are exactly the
\vspace{-.5ex}
\BIT
\IT[{\rm (1$\zeta$)}] $\zeta$-patch spaces of core spaces,
\IT[{\rm (2$\zeta$)}] $\zeta$-convex, core stable (semi-)qospaces,
\IT[{\rm (3$\zeta$)}] $\zeta$-convex, upper regular, locally filtered, ${\uparrow}$-stable and d-stable qospaces.  
\EIT
\ET

\BP
(1) and (2): Let  $T = (X,\leq,\T)$ be a sector space. For $x\in O\in \T$, there are $u \!\in\! X$ and $V \!\in\! \T^{\,\geq}$ such that 
$x\in int_{\T} ({\uparrow\!u} \mathop{\cap} V) = int_{\T} ({\uparrow\!u})\cap V \subseteq {\uparrow\!u} \mathop{\cap} V \subseteq O$.
Then $u\in {\uparrow\!u}\cap V \subseteq O$ and $x \in int_{\T}({\uparrow\!u})$,
whence $O \subseteq \bigcup \,\{ int_{\T}({\uparrow\!u}) : u \!\in\! O\}$.
Using ${\uparrow}$-stability and applying the previous argument to ${\uparrow\!O}$ instead of $O$, one concludes that $T$ is core stable.
Again by ${\uparrow}$-stability, $U = {\uparrow int_{\T}}({\uparrow\!u}\,\cap V)$ belongs to $\T^{\,\leq}$, and the above reasoning yields 
$x\in U\cap V \subseteq {\uparrow\!u}\cap V\subseteq O$, proving strong convexity.

Now, let $(X,\leq,\T)$ be any strongly convex core stable semi-qospace. 
Then, by Lemma \ref{cstable}, $(X,\T^{\leq})$ is a core space with specialization order $\leq$, 
while $\T^{\geq}$ has the dual specialization order. By strong convexity, $\T$ is the patch topology $\T^{\leq} \vee \T^{\geq}$.

On the other hand, by Proposition \ref{pat}, any patch space $(X,\leq, \T )$ of a core (or web) space $(X,\S )$ is a web-quasi-ordered space, in particular ${\uparrow}$-stable.
For $x\in U \cap V$ with $U\in \S$ and $V\in \T^{\geq}$, there are $u\in U$ and $W\in \S$ with 
$x\in W \subseteq {\uparrow\!u}$; it follows that $x\in W \mathop{\cap} V \subseteq {\uparrow\!u} \mathop{\cap} V \subseteq U \mathop{\cap} V$.
Thus,  $(X,\leq, \T )$ is a sector space.

(3) follows from (2) and Lemma \ref{cstable}. Any upper regular semi-qospace is a qospace.

\noindent The modified claims involving coselections $\zeta$ are now easily derived from Proposition\,\,\ref{pat} and Lemma \ref{webpatch}, 
using the fact that core spaces are web spaces.
\EP




We are ready to establish a categorical equivalence between core spaces and $\zeta$-sector spaces. 
As explained in \cite{Epatch}, the right choice of morphisms is a bit delicate.
Continuous maps would be the obvious morphisms between core spaces. 
On the other hand, one would like to have as morphisms between quasi-ordered spaces the isotone continuous maps. 
But, as simple examples in \cite{Epatch} show, a continuous map between two core spaces need not be continuous as a map between the associated $\zeta$-sector spaces,
and a continuous isotone map between $\zeta$-sector spaces need not be continuous for the weak lower topologies (consider the map $f$ on the lattice $L$ in Example \ref{Exnotcon} with
$f(a_n)=f(b_n) =a_0$, $f(c_n)=c_0$). Therefore, we take the $\zeta$-proper maps (see Section \ref{convex}) 
as morphisms between core spaces in order to save the isomorphism theorem. Let us summarize the previous results.

\BPR
\label{sectorcat}
For any coselection $\zeta$, the patch functor ${\rm P}_{\zeta}$ induces a concrete iso\-mor\-phism between 
the category {\bf CS\hspace{.1ex}} (resp.\ {\bf CS}\hspace{.1ex}{\boldmath $\zeta$})
of core spaces with continuous (resp.\ $\zeta$-proper) maps and the category {\boldmath $\zeta$}{\bf S\hspace{.2ex}l} (resp.\ {\boldmath $\zeta$}{\bf S\hspace{.1ex}c})
of $\zeta$-sector spaces with isotone lower semicontinuous (resp.\ continuous) maps. 
The inverse isomorphism is induced by the functor ${\rm U}$, sending a sector space $T$ to the core space ${\rm U}T$.
\EPR

A related class of morphisms is formed by the {\em core continuous maps}, 
i.e.\ continuous maps for which preimages of cores are cores. 
In terms of the specialization orders, the latter condition means that these maps are residual
(preimages of principal filters are principal filters), or equivalently, that they have lower adjoints (cf.\ \cite{Eclo} and \cite[0--1]{CLD}). 
Core continuous maps are $\alpha$-, $\sigma$- and $\upsilon$-proper, but not conversely.
The following facts are established in \cite{EABC} (see also \cite{HM}):

\BCO
\label{qopen}
Passing to lower adjoints yields a duality between the category {\bf CSc} of core spaces 
with core continuous maps and the category {\bf CSq} of core spaces with quasiopen maps (saturations of images of open sets are open) 
that are residuated (preimages of point closures are point closures).
The full subcategory {\bf SCSc} of sober core spaces is equivalent to the category {\bf CDjm} of completely distributive spatial frames 
with maps preserving arbitrary joins and meets, 
which in turn is dual (via adjunction) to the category {\bf CDjw} of completely distributive spatial frames 
with maps preserving joins and the relation $\triangleleft$, where $x \triangleleft\, y \,\Leftrightarrow\,  x\in \bigcap\,\{ \,{\downarrow\!A} : y\leq \bigvee\! A\}$. 
\ECO

\begin{picture}(300,70)
\put(-2,60){\bf CDjm}
\put(-2,2){\bf CDjw}

\put(10,50){\vector(0,-1){30}}
\put(16,20){\vector(0,1){30}}

\put(67,66){\vector(-1,0){30}}
\put(37,60){\vector(1,0){30}}

\put(67,8){\vector(-1,0){30}}
\put(37,2){\vector(1,0){30}}

\put(80,60){\bf SCSc}
\put(80,2){\bf SCSq}

\put(90,50){\vector(0,-1){30}}
\put(96,20){\vector(0,1){30}}

\put(120,61){$\subset$}
\put(123,61){\vector(1,0){30}}

\put(120,3){$\subset$}
\put(123,3){\vector(1,0){30}}

\put(164,60){\bf CSc}
\put(164,2){\bf CSq}

\put(170,50){\vector(0,-1){30}}
\put(176,20){\vector(0,1){30}}

\put(197,61){$\subset$}
\put(220,61){\vector(1,0){10}}
\put(206,60.5){$\dots$}
\put(193,47){\small $\zeta = \alpha,\sigma,\upsilon$}

\put(241,60){{\bf CS}\hspace{.1ex}{\boldmath $\zeta$}}
\put(244,2){{\boldmath $\zeta$}{\bf S\hspace{.1ex}c}}

\put(250,50){\vector(0,-1){30}}
\put(256,20){\vector(0,1){30}}

\put(274,61){$\subset$}
\put(277,61){\vector(1,0){30}}
\put(274,3){$\subset$}
\put(277,3){\vector(1,0){30}}

\put(320,60){\bf CS}
\put(318,2){{\boldmath $\zeta$}{\bf S\hspace{.2ex}l}}

\put(327,50){\vector(0,-1){30}}
\put(333,20){\vector(0,1){30}}
\end{picture}

\vspace{2ex}

In contrast to locally supercompact spaces (core spaces), locally hypercompact spaces need not be $\zeta$-determined, 
and the categorical equivalence between core spaces (i.e.\ locally hypercompact web spaces) and $\zeta$-sector spaces 
does not extend to locally hypercompact spaces without additional restrictions.
The lattice $\widetilde{L}$ in Example \ref{Exnotcon} is locally hypercompact but not locally supercompact in the weak upper topology; in fact, it fails to be a web space: 
the points $b_n$ have no small web neighborhoods. Since $\widetilde{L}$ is not $\zeta$-determined unless $\zeta \widetilde{L} \!=\! \alpha \widetilde{L}$, 
Lemma \ref{pat} does not apply to that case. 
For the complete lattice $M = (X,\leq)$ in Example \ref{Ex33}, the Scott space $\Sigma M$ is locally hypercompact and $\upsilon$-determined, 
but its weak patch space $\Lambda M = {\rm P}_{\upsilon}\Sigma M$ is not ${\uparrow}$-stable: while $\{ a\}$ is open, ${\uparrow\!a}$ is not. 

\newpage

\section{Fan spaces}
\label{fanspaces}

We have seen that, by virtue of the weak patch functor ${\rm P}_{\upsilon}$, the core spaces bijectively correspond
to the $\upsilon$-sector spaces. We shall now give some effective descriptions of these specific qospaces.
By a {\em fan} we mean a nonempty set of the form ${\uparrow\!u}\setminus {\uparrow\!F}$ for some finite $F$.
In case $(X,\leq,\T)$ is an upper semi-qospace, each fan is a sector, in fact, a $\upsilon$-sector. 
An ${\uparrow}$-stable semi-qospace space with a base of fans will be called a {\em fan space}. 
A nonempty product of quasi-ordered spaces is a fan space iff all factors are fan spaces and all but a finite number have a least element.

\BX
\label{Ex72}
{\rm The Euclidean spaces $\Lambda (\mathbb{Q}^n) = (\Lambda \mathbb{Q})^n$ and $\Lambda (\mathbb{R}^n) = (\Lambda \mathbb{R})^n$ are metrizable fan spaces. 
In contrast to $\Lambda (\mathbb{R}^n)$, the rational space $\Lambda (\mathbb{Q}^n)$ is not locally compact, 
whereas the upper spaces $\Sigma (\mathbb{Q}^n)$ and $\Sigma (\mathbb{R}^n)$ are locally super\-compact.
But neither $\Sigma (\mathbb{Q}^n)$ nor $\Sigma (\mathbb{R}^n)$ is sober, since $\mathbb{Q}^n$ and $\mathbb{R}^n$ fail to be up-complete.
}
\EX

\BX
\label{Ex41}
{\rm An open web, a sector, and a fan in the Euclidean plane $\Lambda \,{\mathbb R}^2$

\begin{picture}(300,85)
\put(10,0){
\begin{picture}(100,85)
\multiput(-6,79)(0,-2){22}{.}
\multiput(-6,79)(2,0){10}{.}
\multiput(-6,35)(2,0){18}{.}
\multiput(30,35)(0,-2){18}{.}
\multiput(30,-1)(2,0){22}{.}
\multiput(74,-1)(0,2){10}{.}
\multiput(48,19)(2,0){14}{.}
\multiput(48,19)(0,2){17}{.}
\multiput(48,53)(-2,0){17}{.}
\multiput(14,53)(0,2){14}{.}
\put(39.5,44){\circle*{3}}
\put(5,44){\line(1,0){34}}
\put(4.5,44){\circle{2}}
\put(4.5,45){\line(0,1){28}}
\put(4.5,74){\circle{2}}
\put(39.5,43){\line(0,-1){34}}
\put(39.5,8.5){\circle{2}}
\put(40,8.5){\line(1,0){30}}
\put(70.5,8.5){\circle{2}}

\put(-10,-15){\em open web, not sector}
\end{picture}
}

\put(130,0){
\begin{picture}(100,90)
\put(5,-15){\em sector, not fan}

\put(0,0){\line(0,1){80}}

\put(0,0){\line(1,0){80}}

\multiput(0,79)(2,0){7}{.}
\multiput(14,78)(2,-1){13}{.}
\multiput(40,64)(1.5,-1.5){8}{.}
\multiput(52,51)(1,-2){13}{.}
\multiput(65,25)(1,-1.5){9}{.}
\multiput(74,11)(.7,-2){6}{.}
\end{picture}
}

\put(255,0){
\begin{picture}(100,90)
\put(30,-15){\em fan}

\put(0,0){\line(0,1){78}}
\put(0,0){\line(1,6){13}}
\put(0,0){\line(1,5){13.5}}
\put(0,0){\line(2,5){27}}
\put(0,0){\line(1,2){27.5}}
\put(0,0){\line(3,4){41}}
\put(0,0){\line(1,1){41}}

\put(0,0){\line(4,3){55}}
\put(0,-.3){\line(2,1){55}}
\put(0,-.3){\line(5,2){69}}
\put(0,-.3){\line(5,1){69}}
\put(0,0){\line(6,1){80}}
\put(0,0){\line(1,0){80}}

\multiput(0,78)(2,0){7}{.}
\multiput(12,77)(0,-2){6}{.}
\multiput(12,67)(2,0){7}{.}
\multiput(26,67)(0,-2){7}{.}
\multiput(26,55)(2,0){8}{.}
\multiput(40,55)(0,-2){8}{.}
\multiput(40,41)(2,0){8}{.}
\multiput(54,41)(0,-2){8}{.}
\multiput(54,27)(2,0){8}{.}
\multiput(68,27)(0,-2){8}{.}
\multiput(68,13)(2,0){6}{.}
\multiput(78,13)(0,-2){7}{.}
\end{picture}
}
\end{picture}
}
\EX

\vspace{3.5ex}

In order to see that hyperconvex upper regular semi-qospaces are regular, it suffices to guarantee the separation of points from subbasic closed sets. 
For closed lower sets, this is possible by upper regularity.
For subbasic closed sets ${\uparrow\!x}$ and points $y\in X\setminus {\uparrow\!x}$, 
one finds  an open upper set $U$ and a closed upper set $B$ with ${\uparrow\!x} \subseteq  U \subseteq  B \subseteq  X\setminus{\downarrow\!y}$;
then $X\setminus B$ is an open set disjoint from $U$ and containing\,\,$y$.
Now, from Theorem \ref{sectors} and Proposition \ref{sectorcat}, we infer for the case $\zeta = \upsilon$:

\BT
\label{fans}
The fan spaces are exactly the 
\vspace{-1ex}
\BIT
\IT[{\rm (1$\upsilon$)}] weak patch spaces of core spaces,
\IT[{\rm (2$\upsilon$)}] hyperconvex core stable (semi-)qospaces,
\IT[{\rm (3$\upsilon$)}] hyperconvex, upper regular, locally filtered, up- and d-stable qospaces.  
\EIT
The category of core spaces with continuous (resp.\ $\upsilon$-proper) maps 
is isomorphic to that of fan spaces and lower semicontinuous (resp.\ continuous) isotone maps. 
All fan spaces are not only upper regular but also regular. Moreover, they are uniformizable, and so completely regular in the presence of the Axiom of Choice.
\ET

Let us make the last statement in Theorem \ref{fans} more precise. For the definition of {\em quasi-uniformities} 
and a study of their relationships to (quasi-)ordered spaces, refer to Fletcher and Lindgren \cite{FL} or Nachbin \cite{Na}.
For a quasi-uniformity $\Q$, the dual $\Q^{-1}$ is obtained by exchanging first and second coordinate, and $\Q^*$ is the uniformity generated by $\Q \cup \Q^{-1}$. 
The topology $\tau (\Q)$ generated by $\Q$ consists of all subsets $O$ such that for $x \!\in\! O$ there is a $\,U\!\in\! \Q$ with $xU = \{ y : (x,y) \in U\} \!\subseteq\! O$.
The following results are due to Br\"ummer and K\"unzi \cite{KB} and Lawson \cite{Lss}. 

\BL
\label{localunif}
If $(X,\S )$ is a locally compact space then there is a coarsest quasi-uniformity $\Q$ with $\tau (\Q) = \S$, and $\Q$ is generated by the sets  
 
$\hspace{4ex} U'\!\rightarrow\! U =\{ (x,y)  \in X\times X : x\in U' \Rightarrow y\in U\} = (X\setminus U')\!\times\! X \, \cup \, X\!\times\!U$ 

\noindent with $\,U,U'\in \S$ and $U\ll U'$ in the continuous frame $\S$. Specifically,

{\rm (1)} If $\B$ is a subbase for $\S$ such that $B = \bigcup\,\{ B'\in \B : B'\ll B\}$ for all $B\in \B$, 

\hspace{4ex}then the sets $B'\!\rightarrow\! B$ with $B,B'\in \B$ and $B\ll B'$ generate $\Q$,

{\rm (2)} $\tau (\Q^{-1})$ is the cocompact topology $\tau_{\pi} \S$,

{\rm (3)} $\tau(\Q^*)$ is the patch topology $\S^{\pi}$.
\EL

For core spaces, these conclusions may be strengthened as follows.

\BPR
\label{superunif}
Let $(X,\S )$ be a core space with specialization qoset $Q = (X,\leq )$ and interior relation $\Ro$.\,Then 
there is a coarsest quasi-uniformity $\Q$ with $\tau (\Q) \!= \S$,\,and

{\rm (1)} $\Q$ is generated by the sets $x'\!\Ro \!\rightarrow\! y'\Ro = \{ (x,y) : x' \Ro\, x \Rightarrow  y'\Ro\, y\}$ with $y' \Ro \, x'$,

{\rm (2)} $\tau (Q^{-1})$ is the weak lower topology $\upsilon \widetilde{Q}$,

{\rm (3)} the fan space $(X,\leq, \S^{\upsilon})$ is determined by $\Q$, i.e.\ $\leq \,= \bigcap \Q$ and $\S^{\upsilon} = \tau (\Q^*)$.

\EPR

\BP
\mbox{We apply Lemma \ref{localunif} to the base $\B = \{ x\Ro : x\!\in\! X\}$ of $\O_{\Ro} = \S$ (Theorem \ref{Cspace}).}
By idempotency, $x\,\Ro\, y$ implies $x\,\Ro\, x'\Ro\, y$ for some $x'$, hence $x'\Ro \ll x\Ro$. 
This yields the equation $x\Ro = \bigcup \,\{ x'\!\Ro \in \B : x'\!\Ro \ll x\Ro\}$, and then (1) follows from Lemma \ref{localunif}.

Now, Theorem \ref{Cspace}\,(6) and Lemma \ref{localunif}\,(2) give $\tau (\Q^{-1}) = \tau_{\pi}\S = \upsilon \widetilde{Q}$.
Finally, $\bigcap \Q$ is the specialization order of $\tau (\Q)$, and  $\S^{\upsilon} = \S^{\pi} = \tau (\Q^*)$, by Lemma \ref{localunif}\,(3).

It is also easy to check the equations $\S = \tau (\Q)$ and $\upsilon \widetilde{Q} = \tau (\Q^{-1})$ directly:

For $U = x'\!\Ro \!\rightarrow\! y'\Ro$, the set $xU$ is equal to $y'\Ro$ if $x'\Ro\, x$, and to $X$ otherwise, 

while the set $Uy$ is equal to $X\setminus x'\Ro$ if $y'\!\!\not\!\Ro\, y$, and to $X$ otherwise. 

\noindent In case $y'\Ro\,x'$ but not $y'\Ro\, y$, the inclusion $y\in X\setminus y'\Ro \subseteq X\setminus {\uparrow\!y'} \subseteq x'\Ro = Uy$ 
shows that each $Uy$ is a neighborhood of $y$ in the weak lower topology.
\EP

\vspace{1ex}

In \cite{FL} (see also \cite{Kue}, \cite{Lss}, \cite{Na}), it is shown that in {\sf ZFC} a pospace \mbox{$T = (X,\leq,\T)$} is determined by a quasi-uniformity $\Q$, 
i.e.\ $\leq \ = \bigcap \Q$ and $\T = \tau (\Q^*)$, iff it has a (greatest) order compactification, which in turn is equivalent to saying that
$T$ is {\em completely regular(ly) ordered}, i.e., a pospace such that for $x \!\in\! U \!\in\! \T$, there are continuous $f,g: X \rightarrow [0,1]$ with $f$ isotone and $g$ antitone,
$f(x) = g(x) = 1$, and $f(y) = 0$ or $g(y) = 0$ for $y\in X\setminus U$. Dropping the antisymmetry, we arrive at

\BCO
\label{complreg}
In {\sf ZFC}, every fan space is completely regularly quasi-ordered.
\ECO

\newpage

\section{Continuous domains as pospaces}
\label{domain}

We now are going to establish a representation of continuous domains by certain pospaces, 
generalizing the famous characterization of continuous
lattices as meet-continuous lattices whose Lawson topology is  Hausdorff (see \cite[III--2.11]{CLD}). 
In contrast to the situation of complete lattices, we require at most up-completeness.

For our purposes, we need a further definition. 
Given an upset selection $\zeta$ (see Section \ref{convex}), a quasi-ordered space $(Q,\T) = (X,\leq,\T )$ or its topology is said to be {\em $\zeta$-stable} if
the interior operator $^{\circ}$ of the upper topology $\T^{\leq}$ satisfies the equation
\vspace{-.5ex}
$$
\textstyle{ Y^{\circ} = \bigcup\,\{ ({\uparrow\!y})^{\circ} : y\in Y\}} \ \mbox{ for all } Y \!\in \! \zeta Q. 
$$ 
If $(Q,\T)$ is ${\uparrow}$-stable, the operator $^\circ$ may be replaced by the interior operator of the original topology $\T$.
For instance, {\em core stable} means {\em $\alpha$-stable plus ${\uparrow}$-stable}.\\
We denote by $\curlyvee Q$ and $\curlywedge Q$ the set of all finite unions (incl.\ $\emptyset \!= \bigcup \emptyset$) 
resp.\ intersections (incl.\,$X \!= \bigcap \emptyset$) of  principal filters, i.e., 
the unital join- resp.\ meet-sub\-semi\-lattice they generate in the lattice $\alpha Q$ of all upper sets. 
Further, we denote by $\diamondsuit Q$ the sublattice of $\alpha Q$ generated by the principal filters and containing $\emptyset$ and $X$.

\BL
\label{stable}
Let $T = (Q,\T )$ be any quasi-ordered space.
\vspace{-.5ex}
\BIT
\IT[{\rm (1)}] $T$ is $\curlywedge$-stable iff the sets $\Ro y = \{ x : y\in ({\uparrow\!x})^{\circ}\}$ are directed (ideals). 
\IT[{\rm (2)}] $T$ is $\diamondsuit$-stable iff it is $\curlyvee$-stable and $\curlywedge$-stable.
\IT[{\rm (3)}] $T$ is $\curlyvee$-stable if its upper space ${\rm U}T$ is a web space. 
\IT[{\rm (4)}] $T$ is $\curlywedge$-stable if $Q$ is a join-semilattice with least element. 
\IT[{\rm (5)}] $T$ is $\diamondsuit$-stable if it is a lattice with $0$ and continuous unary $\wedge$-operations.
\IT[{\rm (6)}] $T$ is a $\curlyvee$-stable lower semi-qospace with locally hypercompact upper space ${\rm U}T$
iff the latter is a core space with specialization qoset $Q$.
\EIT
\EL

\BP
(1) Suppose $T$ is $\curlywedge$-stable. For finite $F\!\subseteq\! \Ro y$, the set 
\mbox{$F^{\uparrow\!} = \bigcap \, \{{\uparrow\! x} : x\in F\}$} is a member of $\curlywedge Q$, whence
$y\in \bigcap \, \{ ({\uparrow\! x})^{\circ} : x\in F\} = (F^{\uparrow})^{\circ} = 
\bigcup\,\{ ({\uparrow\! u})^{\circ} : u\!\in\! F^{\uparrow\!}\}$, i.e., $F^{\uparrow}\cap \Ro y \neq \emptyset$. 
In other words, $\Ro y$ is directed (and always a lower set). Conversely, if that is the case then,
for each finite subset $F$ of $Q$, one deduces from $y\in (F^{\uparrow})^{\circ}$ that $F$ is contained in $\Ro y$,
whence there is an upper bound $u$ of $F$ in $\Ro y$; thus, $y\in \bigcup\,\{ ({\uparrow\!u})^{\circ}\! : u \!\in\! F^{\uparrow}\}$.
The reverse inclusion $\bigcup\,\{ ({\uparrow\!u})^{\circ}\! : u \!\in\! F^{\uparrow}\} \!\subseteq\! (F^{\uparrow})^{\circ}$ is clear.

(2) Obviously, $\diamondsuit$-stable quasi-ordered spaces are $\curlyvee$- and $\curlywedge$-stable. Conversely,
suppose $T$ is $\curlyvee$-and $\curlywedge$-stable. Any $Y\in \diamondsuit Q$ is of the form
$Y = \bigcap \,\{ {\uparrow\!F_i} : i \! < \! n\}$ where each of the finitely many sets $F_i$ is finite. Hence,

$Y^{\circ} = \bigcap \,\{ ({\uparrow\!F_i})^{\circ} : i < n \} =$ \hfill { ($\curlyvee$-stability)} 

$\bigcap \,\{ \,\bigcup \,\{ ({\uparrow\!x})^{\circ} : x\in F_i \} : i < n \} =$ \hfill (set distributivity)

$\bigcup \,\{ \,\bigcap\,\{ ({\uparrow\!\chi_i})^{\circ}\! : i \!<\! n \} : \chi \in \prod_{i<n}\! F_i \} =$ \hfill { ($\curlywedge$-stability)} 

$\bigcup\,\{ ({\uparrow\!y})^{\circ}\! : y\in \bigcap\,\{{\uparrow\!\chi_i} : i<n\}, \,\chi \in \prod_{i<n}\! F_i \} =$ \hfill (set distributivity)

$\bigcup\,\{ ({\uparrow\!y})^{\circ} : y\in \bigcap\,\{ {\uparrow\!F_i} : i<n\} = Y \}.$

(3) If ${\rm U}T$ is a web space then its interior operator preserves finite unions \cite{Eweb}; 
in particular, for finite $F$, one has $({\uparrow\!F})^{\circ}\! = \bigcup \,\{ ({\uparrow\!y})^{\circ}\! : y\in F\} = \bigcup \,\{ ({\uparrow\!z})^{\circ}\! : z\in {\uparrow\!F}\}$ 
(as $y\leq z$ implies ${\uparrow\!z}\subseteq {\uparrow\!y}$).

(4) In a join-semilattice $Q$ with least element, $\curlywedge Q$ is the set of all principal filters.

(5) If $Q$ is a lattice with least element and $\T$-continuous unary meet-operations $\wedge_x : y \mapsto x\wedge y$, then ${\rm U}T$ is a web space \cite{Epatch};
hence, $T$ is $\curlyvee$-stable by (3). Furthermore, $T$ is $\curlywedge$-stable by (4), and finally $\diamondsuit$-stable by (2).

(6) Let $T =(X,\leq,\T )$ be a lower semi-qospace. Then $\leq$ is the specialization order of $\T^{\leq}$.
The upper space ${\rm U} T = (X,\T^{\leq})$ is a core space iff  $U \!= \bigcup\,\{ ({\uparrow\!y})^{\circ}\! : y \!\in\! U\}$ for all $U \!\in\! \T^{\leq}$, 
which is equivalent to requiring that $T$ is $\curlyvee$-stable and satisfies the equation 
$U = \bigcup\,\{ ({\uparrow\!F})^{\circ} : F\subseteq U, \, F \mbox{ finite}\}$ for all $U\in \T^{\leq}$.
But the latter condition means that the upper space is locally hypercompact.
\EP

By an {\em mc-ordered space} we mean an ordered space such that every monotone net in the space 
or, equivalently, every directed subset of the space, regarded as a net, has a supremum to which it converges. 
It is shown in \cite{Epatch} that the strongly convex mc-ordered semi-pospaces are exactly the patch spaces of monotone convergence spaces (d-spaces), 
and that the Lawson spaces of domains are just the hyper\-convex, mc-ordered and upper m-determined semi-pospaces, 
where an ordered space is {\em upper m-determined} iff every upper set $U$ intersecting all directed sets whose closure meets $U$ is open.  
We are prepared for the main result in this section, characterizing the Lawson spaces of continuous domains in various ways.

\BT
\label{pospace}
For an ordered space $T$, the following conditions are equivalent:
\vspace{-.5ex}
\BIT
\IT[{\rm (1)}] $T$ is the Lawson space of a continuous domain.
\IT[{\rm (2)}] $T$ is a fan space whose upper space is sober (or a d-space).
\IT[{\rm (3)}] $T$ is the Lawson space of a meet-continuous domain, $\curlywedge$-stable and T$_2$.
\IT[{\rm (4)}] $T$ is a hyperconvex, mc-ordered and core stable (semi-)pospace.
\IT[{\rm (5)}] $T$ is hyperconvex, mc-ordered, ${\uparrow}$-stable, $\diamondsuit$-stable, and T$_2$.
\EIT
\vspace{-.5ex}
Moreover, `T$_2$' may be replaced by `T$_2$-ordered', by 'upper T$_3$-ordered', and by `T$_3$'.
\ET

\BP
(1)$\,\Rightarrow\,$(2): Let $P$ be a continuous domain with $T = \Lambda P$. By Corollary \ref{Cspace}, its upper space $\Sigma P$ is a sober core space,
and by Theorem \ref{fans}, $\Lambda P$ is a fan space. 

(2)$\,\Rightarrow\,$(3): By Theorem\,\ref{fans}, $T =(P,\T )$ is regular, upper regular and T$_1$, i.e., T$_3$ and upper T$_3$-ordered, {\it a fortiori} T$_2$-ordered;
the upper space ${\rm U} T$ is a core space and a d-space.
By Corollary\,\,\ref{Cspace}, $P$ is a continuous, hence meet-continuous domain, $\T^{\leq}$ is its Scott topology $\sigma P$, 
and $\T \!=\! \T^{\leq \upsilon}$ is the Lawson topology $\lambda P$. 
Since ${\rm U} T$ is a core space, $T$ is $\curlywedge$-stable, by Theorem \ref{Cspace}\,(1) and Lemma\,\ref{stable}\,(1). 

(2)$\,\Rightarrow\,$(4): By Theorem \ref{fans}, ordered fan spaces are hyperconvex core stable pospaces, 
and such spaces are mc-ordered if their upper space is a d-space.

(3)$\,\Rightarrow\,$(5):  If $P$ is a meet-continuous domain then $\Sigma P$ is a web space, so 
its weak patch space, the Lawson space $\Lambda P$, is web-ordered, hence ${\uparrow}$-stable (Proposition\,\ref{webpatch}); 
it is a hyperconvex semi-pospace, and mc-ordered since $\Sigma P$ is a d-space.
By Lemma \ref{stable}\,(3), $T$ is $\curlyvee$-stable, and if it is $\curlywedge$-stable, it is $\diamondsuit$-stable by Lemma \ref{stable}\,(2). 

(4)$\,\Rightarrow\,$(5): Core stable semi-pospaces are ${\uparrow}$-stable, $\diamondsuit$-stable and T$_2$ (Lemma\,\ref{cstable}).

(5)$\,\Rightarrow\,$(1): By Lemma \ref{stable}\,(1), the sets $\Ro y = \{ x: y\!\in\! ({\uparrow\! x})^{\circ}\}$ are directed.
As $T$ is mc-ordered, $\Ro y$ has a join $x \!=\! \bigvee \Ro y$. 
Assume $x \!<\! y$ and choose disjoint sets $V,W\in \T$ with $x\in V$ and $y\in W$. 
By hyperconvexity, there is a $U \in \T^{\leq}$ and a finite set $F$ with $x\in U\setminus {\uparrow\! F}\subseteq V$.
Then $y\in U\cap W\subseteq {\uparrow\! F}$ (as $x < y$ and $V\cap W = \emptyset$), 
and $U\cap W\in \T$ yields ${\uparrow\!(U\cap W)}\in \T^{\leq}$ by ${\uparrow}$-stability.
Thus, $y\in ({\uparrow\! F})^{\circ} = \bigcup\,\{ ({\uparrow\! u})^{\circ}\! : u\!\in\! F\}$
by $\curlyvee$-stability, and so $u\in \Ro y$ for some $u\in F$, which leads to the contradiction $u\leq \bigvee \! \Ro y = x\in U\setminus {\uparrow\! F}$. 
Hence, $y$ is the directed join of\,\,$\Ro y$. Since $T =(P,\T )$ is mc-ordered, $\T^{\leq}$ is coarser than $\sigma P$. 
It follows that $\Ro y$ coincides with the way-below ideal $\ll\! y$ (indeed, $x \ll y$ implies $x\,\Ro\,y$, since $\Ro y$ is an ideal with join $y$; 
and $x\,\Ro\, y$ implies $x \!\ll\! y$, since for directed $D$, $x \,\Ro \, y = \bigvee\! D$ entails 
$y\in int_{\T^{\leq}} {\uparrow\!x} \subseteq int_{\sigma P} {\uparrow\!x} \subseteq x\!\!\ll$). 
Therefore, $P$ is continuous, $\T^{\leq} = \sigma P$ (as the base $\{ x\Ro : x\in P\}$ of $\sigma P$ is contained in $\T^{\leq}$), 
and finally, $\T = \T^{\leq \upsilon} = \lambda P$.
\EP

Since all topologies on join-semilattices with 0 are $\curlywedge$-stable, we obtain at once:

\BCO
\label{complete}
A complete\,lattice\,$L$ is continuous iff $\Sigma L$ is a web space and $\Lambda L$\,is\,T$_2$.
In that case, $\Lambda L$ is an (upper) regular pospace (and in {\sf ZFC}, it is even compact).
\ECO

In categorical terminology, parts of Corollary \ref{contdom} and Theorem \ref{pospace} read as follows:

\BPR
\label{iso}
The category {\bf CD} of continuous domains and maps preserving directed joins is concretely isomorphic, 
by virtue of the Scott functor $\Sigma$, to the category {\bf SCS} of sober core spaces, and by virtue of the Lawson functor $\Lambda$,
to the category {\bf MFS} of mc-ordered fan spaces with isotone lower semicontinuous maps. 
The inverse isomorphism $\Lambda^-$ is induced by the forgetful functor $(P,\T) \mapsto P$. 
\EPR

\begin{picture}(300,60)(-80,0)

\put(82,51){{\bf CD}}
\put(35,10){{\bf SCS}}
\put(119,10){{\bf MFS}}
\put(78,45){\vector(-1,-1){20}}
\put(58,35){$\Sigma$}
\put(63,25){\vector(1,1){20}}
\put(77,29){$\Sigma^{\!-}$}
\put(102,45){\vector(1,-1){20}}
\put(117,35){$\Lambda$}
\put(117,25){\vector(-1,1){20}}
\put(93,29){$\Lambda^{\!-}$}
\put(65,11){\vector(1,0){46}}
\put(110,14){\vector(-1,0){46}}
\put(85,17){${\rm U}$}
\put(83,0){${\rm P}_{\upsilon}$}
\end{picture}

\vspace{1ex}

Suitable examples show that the properties in Theorem \ref{pospace}\,(5) are independent; that is, none of them follows from
the combination of the other four properties.

\BX
\label{Ex51}
{\rm
For ${\mathbb I} = \,]\,0,1\,] \subseteq {\mathbb R}$, the left half-open interval topology $\T = \upsilon \hspace{.2ex}{\mathbb I}\,^{\alpha}$ 
makes ${\mathbb I}$ a pospace with all the properties in (5) except hyper- resp.\ strong convexity:
for any interval $]\,0,y\,]$ with $0 \!<\! y \!<\! 1$ and any upper set $U$, there is no lower set 
$V = \,]\,0,\,z\,[$ with $y\in U \cap V \subseteq \ ]\,0,y\,]$, because any $x\in \ ]\,y,z\,[$ belongs to $U\cap V$.
}
\EX

\BX
\label{Ex52}
{\rm
All finite products of chains are continuous posets; but they are domains only if all nonempty subsets have joins. 
The pospaces $\Lambda \,{\mathbb Q}^n$ and $\Lambda \,{\mathbb R}^n$ have all properties in (5) except of being mc-ordered, 
as neither ${\mathbb Q}$ nor ${\mathbb R}$ has a join. 
}
\EX

\BX
\label{Ex53}
{\rm
Consider the subset $X = \{ 0\} \cup \, [ \,1,2\, ] \, \cup \{ 3 \}$ of ${\mathbb R}$, 
equipped with the Euclidean topology inherited from ${\mathbb R}$,
and the order $\sqsubseteq $ defined by 
$$x \sqsubseteq y \ \Leftrightarrow \ x = y \mbox{ or } x = 0 \mbox{ or } y = 1 \mbox{ or } (x = 2 \mbox{ and }  y \in \,]\,1,2\,[\,).$$

\vspace{-1.5ex}

\begin{picture}(200,70)(-130,0)

\put(90,30){\circle{5}}
\put(95,27){$3$}
\put(87.5,32){\line(-3,2){40}}
\put(87.5,28){\line(-3,-2){40}}

\put(45,15){\circle{5}}
\put(45,13){\line(0,-1){10}}

\put(45,60){\circle{5}}
\put(35,58){$1$}
\put(45,0){\circle{5}}
\put(33,34){$]\,1,2\,[$}
\put(35,10){$2$}
\put(35,-4){$0$}

\put(43,17){\line(-2,3){8}}
\put(47,17){\line(2,3){8}}

\put(43,57.5){\line(-2,-3){8}}
\put(47,57.5){\line(2,-3){8}}

\end{picture}

\vspace{2ex}

\noindent This is a compact pospace, hence mc-ordered \cite[VI-1]{CLD} and T$_2$. It is $\curlyvee$-stable, 
since 
$({\uparrow\!F})^{\circ} = {\uparrow (F\cap \{ 0,2\})} \cup (F\cap \{ 3\}) = \bigcup \{ ({\uparrow\!x})^{\circ}: x\in F\}$ for all finite subsets\,\,$F$.
It is $\curlywedge$-stable (because $(X,\sqsubseteq )$ is a complete lattice), and therefore $\diamondsuit$-stable (Lemma \ref{stable}\,(2)). 
Furthermore, it is hyperconvex, in view of the equations

$\{ 0\} = X\setminus {\uparrow\!\{ 2,3\}}$, $\{ 3 \} = (X\setminus \{ 0\} )\setminus {\uparrow\!2}$, and $U = {\uparrow\!U}\setminus {\uparrow\!1}$ for $U \subseteq \ ]1,2\,]$.

\noindent But this pospace fails to be ${\uparrow}$-stable, since $\{ 3\}$ is open, while $\{ 1,3\} = {\uparrow\!3}$ is not. 
}
\EX

\BX
\label{Ex54}
{\rm
If $(X,\T )$ is a nonempty T$_2$-space whose finite subsets have empty interior (e.g.\ $\T = \lambda \,{\mathbb R}$) 
then the pospace $(X, =, \T )$ is trivially up- and $\curlyvee$-stable, 
hyperconvex and mc-ordered, and it is almost $\curlywedge$-stable: for finite subsets $F$ with at least two elements, 
one has $F^{\uparrow} = \emptyset$, hence $(F^{\uparrow})^{\circ} = \bigcup \,\{ ({\uparrow\!y})^{\circ} :y\in F^{\uparrow}\} = \emptyset$.
But, ``by the skin of its teeth'', such a pospace fails to be $\curlywedge$-stable, because for $F = \emptyset$, $F^{\uparrow}$ is the whole ground set $X$,
whence 
$(F^{\uparrow})^{\circ} = X \neq \emptyset = \bigcup \,\{ ({\uparrow\!y})^{\circ} :y\in F^{\uparrow}\! = X\}.$
}
\EX

\BX
\label{Ex55}
{\rm
The completed open real square $S \!= \,]\,0,1\,[^2 \,\mathop{\cup}\, \{ (0,0),(1,1) \}$
endowed with the interval topology is an irreducible, ${\uparrow}$-stable compact semi-pospace, but not T$_2$ (cf.\ Example 3.5 in \cite{Epatch}). 
However, it is hyperconvex, mc-ordered, $\curlyvee$- and $\curlywedge$-stable, and so $\diamondsuit$-stable (Lemma \ref{stable}\,(2)). 
But this space heavily fails to be web-ordered, since the only web neighborhood of any point is the whole space.

}
\EX

\noindent {\bf Note.} The basic results about core spaces and fan spaces are more than 30 years old; they have been reported by the author
at the Annual Meeting of the\,\,DMV, Dortmund 1980\,\cite{Evdt} 
and at the Summer School on Ordered Sets and Universal Algebra, Donovaly 1985
but have not been published in a systematic treatise until now. 
A common theory of web spaces and core spaces is possible by introducing so-called {\em $\kappa$-web spaces}, where $\kappa$ is
a cardinal parameter; that approach was initiated in \cite{Eweb}.
A comprehensive theory of so-called {\em $\zeta$-domains} and their topological manifestation, providing common generalizations of continuous, 
hypercontinuous, algebraic and many other variants of domains or posets, is presented in \cite{Edom}.

\section{Semitopological and topological semilattices}
\label{semi}

Let us now apply some of the previous results to the situation of {\em semilattices}, by which we always mean {\em meet-semilattices}. 
A {\em semilattice-ordered space} is an ordered space $S$ whose underlying poset is a semilattice;
by slight abuse of language, we call it {\em compatible} if the topology is compatible with the semilattice order, and we call a semilattice- and T$_1$-ordered space a T$_1${\em -semilattice}. 
A {\em semitopological semilattice} is a semilattice-ordered\,\,space whose unary meet operations \mbox{$\wedge_x \!: y \mapsto x\wedge y$} are continuous,
while in a {\em topological semilattice} the binary meet operation is continuous.

\vspace{-.5ex}

\BL
\label{top}
A semilattice-ordered space $S =(X,\leq,\T)$ is a topological semilattice whenever one of the following conditions is fulfilled:
\vspace{-.5ex}
\BIT
\IT[{\rm (1)}] $S$ is compatible and locally filtered (a wide web space).
\IT[{\rm (2)}] $S$ has a subbase of sets whose complements are filtered.
\IT[{\rm (3)}] $S$ carries the weak lower topology: $\T = \upsilon (X,\geq)$.
\IT[{\rm (4)}] $S$ is a hyperconvex, locally filtered and ${\uparrow}$-stable semi-pospace.
\EIT
\EL

\vspace{-.5ex}

\BP
(1) For $x,y\in X$ and a filtered neighborhood $D$ of $x\wedge y$ with $D\subseteq U \in \T = \T^{\leq}$, the up-closure $F = {\uparrow\!D}$ is a filter still contained in $U$.
For $W = int_{\T} F$, we obtain $W\wedge W \subseteq F\wedge F \subseteq F \subseteq U$ and $x,y\in W$, as $x\wedge y \in W = {\uparrow\!W}$.

(2) Let $V$ be a subbasic open set such that $F = X\setminus V$ is a filter. If $x,y\in X$ satisfy $x\wedge y\in V$ then $x\in V$ or $y\in V$ (otherwise $x\wedge y \in F$).
Hence, $(x,y)$ lies in $W = (V \!\times\! X)\cup (X \!\times\! V)$, and that is an open set in $S^2$ with $\wedge\, [W] \subseteq V$ 
(indeed, $(u,v)\in W$ implies $u\in V$ or $v\in V$, and so $u\wedge v\in V$, since $V$ is a lower set). 

(3) follows from (2), because $\upsilon (X,\geq )$ has a closed subbase of principal filters.

(4) The ordered space $(X,\leq,\T^{\leq})$ is compatible and locally filtered: for \mbox{$x \!\in\! U \!\in\! \T^{\leq}$,} 
find a filtered $D\subseteq U$ with $x\in W = int_{\T}D$; then ${\uparrow\!D}$ is a filter with $x \!\in\! W \!\subseteq\! {\uparrow\!D} \!\subseteq\! U$,
and ${\uparrow\!W}$ is $\T^{\leq}$-open by ${\uparrow}$-stability.
By (1), the operation $\wedge$ is $\T^{\leq}$-continuous; by (3), it is $\upsilon (X,\geq)$-continuous, 
and then, by hyperconvexity, it is $\T$-continuous.
\EP

We find it convenient to call a topological semilattice {\em s-topological} (resp.\ {\em sc-topological}) if it has small semilattices (resp.\ small convex semilattices),
that is, each point has a neighborhood base consisting of (convex) subsemilattices (cf.\ \cite[VI-3]{CLD}).
We are ready for the characterization of hyperconvex semitopological, resp.\ s-topological T$_1$-semilattices as certain specific web-ordered spaces.

\BT
\label{topsemi}
Let $S$ be a hyperconvex T$_1$-semilattice or, equivalently, the weak patch space of a compatible semilattice-ordered space.

\noindent {\rm (1)} The following three conditions are equivalent:
\vspace{-.5ex}
\BIT
\IT[{\rm (w11)}] $S$ is the weak patch space of a web space.
\IT[{\rm (w12)}] $S$ is a web-ordered space. 
\IT[{\rm (w13)}] $S$ is a semitopological semilattice.
\EIT
\noindent {\rm (2)} The following three conditions are equivalent:
\vspace{-.5ex}
\BIT
\IT[{\rm (w21)}] $S$ is the weak patch space of a wide web space.
\IT[{\rm (w22)}] $S$ is a locally filtered and ${\uparrow}$-stable ordered space. 
\IT[{\rm (w23)}] $S$ is an s(c)-topological semilattice.
\EIT
\noindent {\rm (3)} The following three conditions are equivalent:
\vspace{-.5ex}
\BIT
\IT[{\rm (w31)}] $S$ is the weak patch space of a worldwide web space (a core space).
\IT[{\rm (w32)}] $S$ is a core stable pospace (a fan space). 
\IT[{\rm (w33)}] $S$ is an s(c)-topological semilattice, and ${\rm U}S$ is  locally compact.
\EIT
\ET

\BP
Notice that hyperconvexity ensures closedness of all principal filters, so it suffices to assume that $S$ is a hyperconvex semilattice-ordered lower semi-pospace.

(w11)$\,\Leftrightarrow \,$(w12): Apply Proposition \ref{webpatch} to $\zeta = \upsilon$. 

(w12)$\,\Rightarrow \,$(w13): By Proposition \ref{webpatch}, the upper space $(X,\T^{\leq})$ of $S = (X,\leq, \T)$ is a web space, and $S$ is its weak patch space.
Therefore, as remarked in \cite{Eweb} and \cite{Epatch}, the unary meet operations $\wedge_x$ are $\T^{\leq}$-continuous;
by Lemma \ref{top}\,(3), they are $\upsilon (X,\geq)$-continuous, and by hyperconvexity, they are also $\T$-continuous.

(w13)$\,\Rightarrow \,$(w12): This was shown in \cite{Eweb}.

(w21)$\,\Leftrightarrow \,$(w22): Apply Theorem \ref{wideweb} to $\zeta = \upsilon$.

(w22)$\,\Rightarrow \,$(w23): By Lemma \ref{top}\,(4), $S$ is a topological semilattice.
Given $x \!\in\! O \!\in\! \T$, use hyperconvexity and local filteredness in order to find a $U\in \T^{\leq}$, a finite set $F$, 
and a filtered set $D$ such that $x\in int_{\T} D \subseteq D \subseteq U \setminus {\uparrow\!F} \subseteq O$.
Then ${\uparrow\!D}\setminus {\uparrow\!F}$ is a convex subsemilattice with 
$x\in int_{\T} D \subseteq {\uparrow\!D}\setminus {\uparrow\!F} \subseteq U\setminus {\uparrow\!F} \subseteq O$,
which shows that $S$ has small convex semilattices.

(w23)$\,\Rightarrow \,$(w22): By continuity of the unary operations $\wedge_x$, $S$ is ${\uparrow}$-stable: indeed,
$O\in \T$ implies ${\uparrow\!O} = \bigcup\,\{ \wedge_x^{-1}[O] : x\in O\} \in \T$. Clearly, subsemilattices are filtered.

The equivalence of (w31), (w32) and (w33) is now easily verified with the help of (2), Theorem \ref{fans} 
and Proposition \ref{supertop}, which says, among other things, that the core spaces are exactly the locally compact web spaces  (cf.\ \cite[VI--3.3]{CLD}).
\EP

\vspace{1ex}

Combining Theorem \ref{pospace} with Theorem \ref{topsemi}, we arrive at

\BCO
\label{contsemi}
For a semilattice-ordered space $S$, the following are equivalent:
\vspace{-.5ex}
\BIT
\IT[{\rm (w41)}] $S$ is the weak patch space of a (unique) sober core space.
\IT[{\rm (w42)}] $S$ is an mc-ordered fan space. 
\IT[{\rm (w43)}] $S$ is a hyperconvex, mc-ordered, s-topological T$_1$-semilattice with locally compact upper space ${\rm U}S$.
\IT[{\rm (w44)}] $S$ is the Lawson space of a continuous semilattice.
\EIT
The weak patch functor ${\rm P}_{\upsilon}$ induces concrete isomorphisms between
\vspace{-.5ex}
\BIT
\IT[{\rm (1)}] the category of compatible semitopological semilattices,\,i.e.\,semilattice-ordered web spaces, 
and that of hyperconvex semitopological T$_1$-semilattices,
\IT[{\rm (2)}] the category of compatible s-topological semilattices,\,i.e.\,semilattice-ordered wide web spaces, 
and that of hyperconvex s-topological T$_1$-semilattices,
\IT[{\rm (3)}] the category of locally compact compatible s-topological semilattices,\,i.e.\,semi\-lattice-ordered core spaces, 
and that of topological semilattice fan spaces,
\IT[{\rm (4)}] the category of Scott\,spaces of continuous\,semilattices,\,i.e.\ semilattice-ordered sober core spaces, 
and that of Lawson spaces of continuous\,semilattices,\,i.e.\ mc-ordered topological semilattices that are fan spaces.
\EIT
\ECO

\noindent Of course, most elegant results are available for compact pospaces. From \cite[IV--1]{FL} or \cite[VI--1]{CLD}, we learn the following facts:
every compact pospace is
\vspace{-.5ex}
\BIT
\IT[--] mc-ordered and dually mc-ordered,
\IT[--] monotone normal, in particular upper and lower regular,
\IT[--] strongly convex.
\EIT
For any closed subset $C$ of a compact pospace $T$, the sets ${\uparrow\!C}$ and ${\downarrow\!C}$ are closed, 
but $T$ need neither be ${\uparrow}$-stable nor locally filtered: see Examples \ref{Ex33} and \ref{Ex53}.

A semilattice is said to be {\em complete} if all directed subsets have suprema and all nonempty subsets have infima
(this together with the existence of a top element defines a complete lattice).
Recall from  \cite[IV--3]{CLD} the famous 

\vspace{1ex}

\noindent {\bf Fundamental Theorem of Compact Semilattices}\\
{\em The Lawson spaces of complete continuous semilattices are exactly the compact T$_2$ s-topological semilattices.
}

\vspace{1ex}

This is now a rather easy consequence of the previous facts, the Ultrafilter Theorem (giving compactness of $\Lambda L$ \cite{EPS}) and the following ``non-complete'' version:

\BPR
\label{compact}
The compact Lawson spaces of continuous domains are exactly the locally filtered, ${\uparrow}$-stable compact pospaces.
\EPR

\BP
If $P$ is a continuous domain then, by Theorems \ref{fans} and \ref{pospace}, the Lawson space $\Lambda P$ is a fan space, hence a locally filtered ${\uparrow}$-stable pospace.
Conversely, if $T = (X,\leq,\T)$ is any locally filtered, ${\uparrow}$-stable compact pospace then $T$ is upper regular, mc-ordered and dually mc-ordered; by Lemma \ref{cstable}, it is d-stable,
and its upper space is a core space and a d-space (because $T$ is mc-ordered); hence, by Corollary \ref{contdom}, it is the Scott space of the continuous domain $P =(X,\leq)$.
It follows that the Lawson topology $\lambda P = \sigma P \vee \upsilon \widetilde{P}$ is contained in $\T = \T^{\leq}\vee \T^{\geq}$ (as $T$ is strongly convex).
Now, $\lambda P$ is T$_2$ and $\T$ is compact, so both topologies must coincide. 
\EP

In \cite[III--5]{CLD}, one finds a whole collection of various equivalent characterizations of continuous domains that are compact in their Lawson topology. 
For a detailed study of joint and separate continuity of operations in posets and lattices, see \cite{EG}. 

\newpage

\section{Domain bases and core bases}

Following \cite{CLD}, we mean by a {\em basis} of a domain $P$ a subset $B$ such that for each $y\in P$, the set $\{ b \in\! B: b \ll y\}$ is directed with join $y$.
The pair $(P,B)$ is then referred to as a {\em based domain}.
Note that $P$ is continuous iff it has at least one basis. 

By a {\em core basis} for a space $(X,\S)$, we mean a subset $B$ of $X$ such that for all $U\in \S$ and all $y\in U$,
the set $\Ro_{\S}y$ meets $B\cap U$; i.e., $y\in int_{\S}({\uparrow\!x}) \subseteq {\uparrow\!x} \subseteq U$ for some $x\in B$;
in other words, all points have neighborhood bases formed by cores of elements of $B$.
(To avoid ambiguities, we use the word {\em basis} for subsets of $X$ and the word {\em base} for subsets of the power set of $X$.)
Thus, a space is a core space iff it has a core basis. By a {\em core based space} we mean a pair consisting of a (core) space and a core basis of it.
The following extension of Corollary \ref{contdom} is straightforward:

\BL
\label{basis}
The bases of a domain are the core bases of its Scott space. 
Hence, via the Scott functor $\Sigma$, the based domains correspond to the core based sober spaces. 
\EL

\BPR
\label{Cbasis}
The C-ordered sets are exactly the pairs $(B,\ll\!\!|_B)$ where $B$ is a basis of a domain (which is uniquely determind up to isomorphism by $(B,\ll\!\!|_B)$). 
\EPR

\BP
That, for any basis $B$ of a domain, the pair $(B,\ll\!\!|_B)$ is a C-ordered set follows easily from the interpolation property of $\ll$ (cf.\ Lemma \ref{conti}).
Conversely, any C-ordered set $(X,\Ro )$ is isomorphic to $(\B_{\Ro}, \ll\!\!|_{\B_{\Ro}})$,
where $\B_{\Ro} = \{ \Ro x : x\in X\}$ is a basis of the continuous domain $\I_{\Ro}$ of rounded ideals (see Theorem \ref{Cspace}\,(4)). 
\EP

\BCO
\label{Cor}
The T$_0$ core spaces are exactly the core bases of sober core spaces, equipped with the induced topology. 
\ECO

On the pointfree side, we define a {\em based supercontinuous lattice} to be a pair consisting of a supercontinuous 
(i.e.\ completely distributive) lattice $L$ and a {\em (coprime) basis} of $L$, that is, a join-dense subset of coprime elements.

We are ready for six different descriptions of C-ordered sets:

\BT
\label{Cequiv}
The following six categories are mutually equivalent:

\vspace{1ex}

\begin{tabular}{|l|l|}
\hline
objects & morphisms\\
\hline
\hline
C-ordered sets & interpolating isotone maps\\
\hline
T$_0$ core spaces & continuous maps\\
\hline
fan ordered spaces & lower semicontinuous isotone maps\\
\hline
based domains & maps preserving the bases and directed joins\\
\hline
core based sober spaces & continuous maps preserving the core bases\\
\hline
based supercontinuous lattices & maps preserving the bases and joins\\
\hline
\end{tabular}
\ET

\vspace{1ex}

\noindent On the object level, these equivalences easily follow from Theorem \ref{Cspace}, Corollary\,\ref{contdom}, Theorem \ref{fans}, 
Lemma \ref{Cbasis}, Proposition \ref{Cequiv} and Corollary \ref{Cor}.
The verification of the claimed correspondences between the morphisms is left as an exercise.

\newpage

\section{Density and weight}
\label{Cbases}

We now are looking for characterizations of core bases in terms of the interior relation.
Recall that the lower quasi-order $\leq_{\Ro}$ of an arbitrary relation $\Ro$ on a set $X$ is given by $x \leq_{\Ro} y \,\Leftrightarrow\, \Ro x \subseteq \Ro y$.
Now, we say a subset $B$ of $X$ is 
\BIT
\IT[--] {\em $\Ro$-dense} if $x\,\Ro\,y$ implies $x\,\Ro\,b$ and $b\,\Ro\,y$ for some $b\in B$, 
\IT[--] {\em $\Ro$-cofinal} if $x\,\Ro\,y$ implies $x\!\leq_{\Ro}\!b$ and $b\,\Ro\,y$ for some $b\in B$. 
\EIT

It is straightforward to see that $B$ is $\Ro$-dense and $\Ro$ is transitive iff $B$ is $\Ro$-cofinal and $\Ro$ is idempotent.


Note that the {\em strong patch topology} $\S^{\alpha} = \S \vee \alpha (X,\geq)$ of a space $(X,\S )$ concides with the {\em Skula topology} generated by $\S \cup \S^c$ \cite{Sk}. 
Hence, the sets $C\setminus D$ with $C,D\in \S^c$ form a base for $\S^{\alpha}$.

\BPR
\label{corebase}
Let $(X,\S)$ be a core space and $\Ro$ the corresponding interior relation.
For a subset $B$ of $X$, the following conditions are equivalent:
\BIT
\IT[{\rm (1)}] $B$ is $\Ro$-dense in $X$.
\IT[{\rm (2)}] $B$ is $\Ro$-cofinal in $X$.
\IT[{\rm (3)}] $B$ is a core basis for $(X,\S)$.
\IT[{\rm (4)}] $B$ is dense in $(X,\S^{\alpha})$ and so in any patch space of $(X,\S )$. 
\IT[{\rm (5)}] $\{ {\downarrow\!b} : b\in B\}$ is join-dense in the coframe $\S^c$ of closed sets.
\EIT
\EPR

\BP
(1)$\,\Leftrightarrow\,$(2): The above remark applies, as $\Ro$ is idempotent by Theorem \ref{Cspace}\,(1). 

(1)$\,\Rightarrow\,$(3): $y\in U\in \S$ means $U = U\Ro$ and $x\,\Ro\,y$ for some $x\in U$. 
Choose $b\in B$ with $x\,\Ro\, b \, \Ro \, y$.
Then, $b\in \Ro y$ and, by transitivity,  $b\in x\Ro \subseteq x\!\leq_{\Ro} \ = {\uparrow\!x} \subseteq U$.

(3)$\,\Rightarrow\,$(4): If $C,D$ are $\S$-closed sets with $C \!\not\subseteq\! D$, pick an $x\in C\setminus D$ 
and find a $b\in B$ with $x\in int_{\S}({\uparrow\!b}) \subseteq {\uparrow\!b} \subseteq X\setminus D$. It follows that $b\in C$,
since $b\leq x \in C = {\downarrow\!C}$. Thus, $b\in B\cap (C\setminus D)$. This proves density of $B$ in $\S^{\alpha} = \S \vee \S^c$. 

(4)$\,\Leftrightarrow\,$(5) holds for arbitrary spaces. Recall that ${\downarrow\!x}$ is the closure of the singleton $\{ x\}$ in $(X,\S)$. 
Now, \mbox{$\{ {\downarrow\!b} : b\in B\}$} is join-dense in $\S^c$
iff for $C,D\in \S^c$ with $C\not\subseteq D$ there is a $b\in B$ with ${\downarrow\!b}\subseteq C$ 
but not ${\downarrow\!b}\subseteq D$, i.e.\ $b\in C\setminus D$ (as $C$ and $D$ are lower sets).
But the latter means that $B$ meets every nonempty open set in $\S^{\alpha} = \S \vee \S^c$.

(4)$\,\Rightarrow\,$(1): Suppose $x\,\Ro\, y$ and choose a $z$ with $x\,\Ro\, z \, \Ro\, y$. 
Then $x\Ro \in \O_{\Ro} = \S$ and therefore $z \in x\Ro \cap {\downarrow\!z} \in \S^{\alpha}$. 
Hence, there is a $b \in B \cap x\Ro \cap {\downarrow\!z}$,
and it follows that $x \, \Ro \, b \, \Ro \, y$, since $\Ro y$ is a lower set containing $z$ and so $b$. 
\EP

From now on, we assume the validity of the Axiom of Choice. Hence, each set $X$ has a cardinality, 
represented by the smallest ordinal number equipotent to\,\,$X$. 
The {\em weight} $w(\S)$ resp.\ {\em density} $d(\S)$ of a space or its topology $\S$ is the least possible cardinality of bases resp.\ dense subsets.
The weight of a core space is at most the cardinality of any core basis $B$, since $B$ gives rise to a base $\{ b\Ro = int_{\S}({\uparrow\!b}) : b \in B\}$. 

\BL
\label{cardbase}
Every core basis of a core space $(X,\S)$ contains a core basis of cardinality $w(\S)$. Hence, 
$w(\S)$ is the minimal cardinality of core bases for $(X,\S)$.
\EL

\BP
Let $\B$ be an arbitrary base and $B$ a core basis for $(X,\S)$. The Axiom of Choice 
gives a function picking an element $b_{\,U,V}$\,from each nonempty set of the form
$$
B_{\,U,V} = \{ b \in B : U \subseteq {\uparrow\!b} \subseteq V \} \ \ (U,V\in \B ).
$$
Then
$
B_0 = \{ b_{\,U,V} : U,V\in \B, \ B_{\,U,V} \not = \emptyset\}
$
is a subset of $B$ and still a core basis; in fact, $x\in V \in \B$ implies
\mbox{$x\in U \subseteq {\uparrow\!b} \subseteq V$} for suitable $b\in B$ and $U\in \B$, and it follows that 
$x \in U \subseteq {\uparrow\!b_{\,U,V}} \subseteq V$.
If $\S$ is infinite then so is $\B$, and consequently $|B_0| \leq |\B|^2 = |\B|$. Thus, we get $|B_0| \leq w(\S)$, 
and the remark before yields equality.

If $\S$ is finite then the cores form the least base $\{ {\uparrow\!x} : x\in X\}$,
and choosing a set of representatives from these cores, one obtains a core basis of cardinality $w(\S)$.
\EP

\BX
\label{Ex62}
{\rm
Consider the ordinal space $C = \omega \!+\! 2 = \{ 0, 1, ..., \omega, \omega\!+\!1\}$ with the upper (Scott) topology.
While $\omega \cup \{ \omega\!+\!1\} = C\setminus \{ \omega \}$ is a core basis, the set $B = \omega\!+\!1 = \omega \cup \{ \omega\}$ is {\em not} a core basis:
$U \!= \{ \omega\!+\!1\}$ is $\upsilon$-open but disjoint from $B$.
Nevertheless, $\upsilon C \setminus \{ \emptyset\} = \{ b\Ro : b \in B\}$ is a base (note $x\Ro ={\uparrow\!x}$ for $x\neq  \omega$).
}
\EX
Generalizing the weight of topologies, one defines the {\em weight} $w(L)$ of a lattice $L$ as the least possible cardinality of join-dense subsets of $L$.
For any relation $\Ro$ on a set $X$, we define the {\em $\Ro$-cofinality}, denoted by $c(X,\Ro)$, 
to be the minimal cardinality of $\Ro$-cofinal subsets of $X$. If $\Ro$ is idempotent then $c(X,\Ro)$
is also the {\em $\Ro$-density}, the least cardinality of $\Ro$-dense subsets.
From Proposition \ref{corebase} and Lemma \ref{cardbase} we infer:

\BT
\label{w}
For a core space $(X,\S)$ with interior relation $\Ro$ and any patch topology $\T$ of $\S$, 
the density of $(X,\T)$ is equal to each of the following cardinal invariants:
\vspace{-.5ex}
$$
c(X,\Ro) = w(\S) = w(\S^c) = w(\S^{\upsilon}). 
$$
\ET

Indeed, if $B$ is a core basis for $(X,\S)$ then $\B = \{ bR \setminus {\uparrow\!F} : b\in B,\, F \!\subseteq\! B, \, F \,\mbox{finite}\}$ is a base for $\S^{\upsilon}$,
and if $B$ is infinite, $\B$ and $B$ have the same cardinality. If $B$ is finite, \mbox{the base $\{ {\uparrow\!b} : b \in B\}$ of $\S$ 
is equipotent to the base $\{ {\uparrow\!b}\cap {\downarrow\!b} : b \in B\}$ of $\S^{\upsilon} = \S^{\alpha}$.} 

Since in {\sf ZCF} any supercontinuous lattice is isomorphic to the topology of a core space \cite{EABC}, 
Theorem \ref{w} entails a fact that was shown choice-freely in \cite{ECD}:

\BCO
The weight of a supercontinuous (i.e.\ completely distributive) lattice is equal to the weight of the dual lattice. 
\ECO

\BX
\label{Ex61}
{\rm
On the real line with the upper (Scott) topology $\S$, the interior relation is the usual $<\,$. 
The rationals form a core basis ${\mathbb Q}$, being $<$-dense in ${\mathbb R}$. Hence,
\vspace{-1ex}
$$\omega = c({\mathbb R},<) = w(\S) = w(\S^c) = w(\S^{\upsilon}) \neq w(\S^{\alpha})
$$
because the half-open interval topology $\S^{\alpha}$ has no countable base.
}
\EX

\end{document}